# EURETILE 2010-2012 summary: first three years of activity of the European Reference Tiled Experiment.

Pier Stanislao Paolucci, Iuliana Bacivarov, Gert Goossens, Rainer Leupers, Frédéric Rousseau, Christoph Schumacher, Lothar Thiele, Piero Vicini

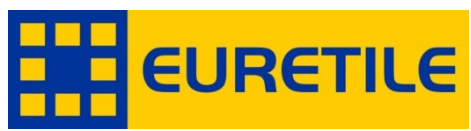

www.euretile.eu

Project: **EURETILE** – European Reference Tiled Architecture Experiment
Grant Agreement no.: **247846**
Call: FP7-ICT-2009-4 Objective: FET - ICT-2009.8.1 Concurrent Tera-device Computing

**ACKNOWLEDGMENT**
The authors summarized in this document the work performed by the EURETILE team during the first three years of the project.
During the period 2010-2012 the full team was composed by:

**Istituto Nazionale di Fisica Nucleare - sezione di Roma (INFN)**
Roberto Ammendola, Andrea Biagioni, Ottorino Frezza, Michela Giovagnoli, Francesca Lo Cicero, Alessandro Lonardo, Pier Stanislao Paolucci, Francesco Simula, Davide Rossetti, Laura Tosoratto, Piero Vicini

**RWTH Aachen University – ISS & SSS (RWTH)**
Jovana Jovic, Rainer Leupers, Luis Murillo, Christoph Schumacher, Jan Henrik Weinstock

**ETH Zurich - The Swiss Federal Institute of Technology Zurich (ETHZ)**
Iuliana Bacivarov, Devendra Rai, Lars Schor, Lothar Thiele, Hoeseok Yang

**Université Joseph Fourier – Grenoble Institute of Technology  - TIMA Laboratories (UJF/INP TIMA)**
Ashraf El Antably, Ikbel Belaid, Clément Deschamps, Nicolas Fournel, Mohamad Jaber, Julian Michaud, Etienne Ripert, Frédéric Rousseau

**Target Compiler Technologies (TARGET)**
Gert Goossens, Werner Geurts

This publication is a derivative of the EURETILE 2010-2012 Publishable Executive Summary. All project deliverable documents have been peer reviewed by a committee of four experts of the field.

The EURETILE project is funded by the European Commission, through the Grant Agreement no. 247846, Call: FP7-ICT-2009-4 Objective FET-ICT-2009.8.1 Concurrent Tera-device Computing.

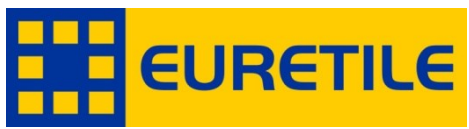

www.euretile.eu

# Table of Contents

# 1. Abstract

This is the summary of first three years of activity of the EURETILE FP7 project 247846. EURETILE investigates and implements brain-inspired and fault-tolerant foundational innovations to the system architecture of massively parallel tiled computer architectures and the corresponding programming paradigm. The execution targets are a many-tile HW platform, and a many-tile simulator. A set of SW process - HW tile mapping candidates is generated by the holistic SW tool-chain using a combination of analytic and bio-inspired methods. The Hardware dependent Software is then generated, providing OS services with maximum efficiency/minimal overhead. The many-tile simulator collects profiling data, closing the loop of the SW tool chain. Fine-grain parallelism inside processes is exploited by optimized intra-tile compilation techniques, but the project focus is above the level of the elementary tile. The elementary HW tile is a multi-processor, which includes a fault tolerant Distributed Network Processor (for inter-tile communication) and ASIP accelerators. Furthermore, EURETILE investigates and implements the innovations for equipping the elementary HW tile with high-bandwidth, low-latency brain-like inter-tile communication emulating 3 levels of connection hierarchy, namely neural columns, cortical areas and cortex, and develops a dedicated cortical simulation benchmark: DPSNN-STDP (Distributed Polychronous Spiking Neural Net with synaptic Spiking Time Dependent Plasticity). EURETILE leverages on the multi-tile HW paradigm and SW tool-chain developed by the FET-ACA SHAPES Integrated Project (2006-2009).

The APE Parallel Computing Lab of INFN Roma is in charge of the EURETILE HW Design (QUonG system/APENet+ board/DNP (Distributed Network Processor) and Scientific Application Benchmarks. The Computer Engineering and Networks Laboratory (TIK) of ETH Zurich (Swiss Federal Institute of Technology) designs the high-level explicit parallel programming and automatic mapping tool (DOL/DAL) and a set of "Embedded Systems" benchmarks. The Software for Systems on Silicon (SSS) of the ISS institute of RWTH Aachen, investigates and provides the parallel simulation technology and scalable simulation-based profiling/debugging support. The TIMA Laboratory of the University Joseph Fourier in Grenoble explores and deploys the HdS (Hardware dependent Software) including the distributed OS architecture. TARGET Compiler Technologies, the Belgian leading provider of retargetable software tools and compilers for the design, programming, and verification of application-specific processors (ASIPs), is in charge of the HW/SW Co-design tools for custom components of the EURETILE architecture.

Grant Agreement no. 247846 Call: FP7-ICT-2009-4 Objective FET-ICT-2009.8.1 Concurrent Tera-device Computing

Scientific Coordinator: Pier Stanislao Paolucci, Istituto Nazionale di Fisica Nucleare, Roma, Italy
Administrative Coordinator: Michela Giovagnoli, Istituto Nazionale di Fisica Nucleare, Roma, Italy

## 1.1. Participating Institutions

- INFN (Istituto Nazionale di Fisica Nucleare) Sezione di Roma (coordinator): the APE Parallel Computing Lab;
- The Computer Engineering and Networks Laboratory (TIK) of ETH Zurich (Swiss Federal Institute of Technology);
- The Software for Systems on Silicon (SSS) of the ISS institute of RWTH Aachen;
- UJF-TIMA: the TIMA Laboratory of the University Joseph Fourier in Grenoble;
- TARGET Compiler Technologies

# 2. Structure of the Document

This document summarizes the activities of the European Reference Tiled Experiment, the EURETILE FET (Future Emerging Technologies) FP7 project 247836, during its first three years of activity (2010-2012).

The document should provide a global view of the project framework, and it should simplify the reading of individual publications produced during the project time frame.

Another purpose is to create an historical record of the project development flow. Therefore, individual sections are dedicated to each year of activity, and they describe the project status as reported at the end of 2010, 2011 and 2012.

As the EURETILE project is an evolution of the SHAPES FP7 project, we inserted also a short summary of the SHAPES results that has been relevant for the EURETILE project.

## 2.1. Bibliographical Notes

The **"References"** section at the end of the document is composed of four groups of citations.

The first group contains a restricted set of "prior art" publications produced by other groups before 2010, the start date of the EURETILE project. Without any claim for completeness, we list a few publications that surely contributed to define the starting point of the project.

The second group contains a selection of publications produced by members of our team before the start of the project, mainly during the time frame of the SHAPES project (2006-2009).

The third group is a selection of publications produced by other groups during the 2010-2012 period about topics that are similar to those investigated by the EURETILE project.

The last set is a selection of the publications produced by the members of the EURETILE team as a result of the project activities. The complete list of presentations and publication produced by the project is included in the **"Dissemination"** section.

The software design for specific platforms follows the methodologies described in the reference paper from Sangiovanni-Vincentelli, Martin (2001). The software is generated from a flow that respects the Y-chart approach proposed by Kienhuis and all (2002), taking into account the architecture specification. It includes few layers, and the lowest layer is called the Hardware dependant Software (HdS) as it is related to hardware access from the software. HdS development process is described in Schirner , Domer and Gerstlauer (2009).

The Kahn process networks model of computation (Khan 1974) is used to express the computational kernels that constitute the lower levels of hierarchy of EURETILE applications, leveraging on the Distributed Operating Layer (DOL) (Thiele, Bacivarov, Haid, Huang, 2007). developed in the framework of the SHAPES project (Paolucci, Jerraya, Leupers, Thiele, Vicini., 2006). The KPN model explicitly separates computation and communication in the parallel application specification: an application kernel is expressed as a set of concurrently executing processes that only communicate via point-to-point FIFO channels. In EURETILE, the individual KPN processes are specified in C/C++, while the topology of the KPN is described using XML-based formats (Haid et al. 2009), that specify the instantiation of processes, their connection by FIFO channels, and their mapping onto the target platforms. This way, the processes that compose the KPN can be compiled for any processor for which a C/C++ compiler is available. Moreover, the specification of the mapping in an XML file rather than in the source code of processes makes the description of the application kernel more platform-independent. The expression of dynamic application scenarios and the specification of upper level of application hierarchies are among the targets addressed by the development of the EURETILE Distributed Application Layer (DAL) (see section 3.1.1).

In the multi-tile context, one challenge is to manage as much efficiently as possible the communication between tiles. We focus the work in EURETILE on the software driver development for HdS generation, while some other works focus on the mapping quality (Lin, Gerstlauer, Evan, 2012) in order to reduce the impact on performances. Our software driver generation depends highly on the hardware architecture, which is a costly approach when dealing with many-tile system with several kinds of tiles. One way to facilitate the re-use of existing drivers is probably the use of virtualization environment (Heiser, 2011). Such a virtualization environment may help as well for multi-application support.

The simulation of the EURETILE many-tile system is a key aspect of the project. The work "Exploiting parallelism and structure to accelerate the simulation of chip multi-processors" experiments with a custom parallel simulation framework (Penry et al., 2006). It discusses the impact of varying system sizes and accompanying cache effects on the simulation host machine, and suggests taking these cache effects into account for the scheduling of logical processes to simulation host cores. Ezudheen et al (2009) "Parallelizing SystemC kernel for fast hardware simulation on SMP machines" is one of the early works describing parallel SystemC simulation that relies on parallel execution of process executions within a delta-cycle. Various methods to schedule SystemC logical processes to physical simulation host OS threads are discussed. Lu et al. (2008) "Learning from mistakes: A comprehensive study on real world concurrency bug characteristics" presented an exhaustive analysis and classification of concurrency bugs (e.g., atomicity violations, order violations) observed in mature real-world parallel applications, as well as a set of conclusions and facts which set a precedent and a guide for research on modern debug methodologies and tools for parallel systems.

One of the goals of EURETILE is to develop the DPSNN-STDP benchmark (Distributed Polychronous Spiking Neural Net with synaptic Spiking Time Dependent Plasticity). The recent paper "Cognitive Computing" (Modha et al., 2011) optimally describes the boundaries of the arena where we intend to play. We resume here a few key references that contributed to define the starting-point of our work. Song (2000) "Competitive Hebbian learning through spike-timing-dependent synaptic plasticity" described the importance of precise time ordering between pre and post-synaptic spiking as a way to capture causal/anti-causal relations between pairs of neurons and decide if to potentiate or depress an individual synapse. In our opinion, this is one of the key features to be captured if brain-inspired model of computations have to be applied to real-world problems. The importance of polychronization, i.e. the relevance of individual axonal delays on computation, is described by Izhikevich (2006). This is another feature that we deemed essential in our cortical simulation benchmark, to create an interconnection topology that could inspire advanced models of hardware interconnection and brain-inspired computational models. We followed Izhikevich (2003, 2004, 2006) also for what concern the model of the spiking neuron. We surely acknowledge the importance of the SpiNNaker Massively-Parallel Neural Net Simulator (Steve Furber et al. 2008) as source of thought about the characteristics of a brain-inspired hardware interconnection systems.

As the EURETILE project is a continuation of the SHAPES project (2006-2009), we listed a few key references about the SHAPES hardware and software tool-chain architecture (Thiele et al, 2006, 2007; Paolucci et al. 2006, Rousseau, Jerraya et al. 2008). The simulation techniques developed in the SHAPES time-framework are summarized by a few papers (Kraemer, Leupers et al., 2007; Schumacher, Leupers et al. 2010). A key role in the hardware architecture of the SHAPES tile has been played by Diopsis MPSOC (Paolucci et al. 2003, 2006) architecture and by the mAgic VLIW DSP (Paolucci et al. 2001, 2002, 2004, 2008).

# 3. Publishable summary – 2010-2012 revision

## 3.1. Project Objectives and Expected Final Results

The EURETILE project (**http://www.euretile.eu**) investigates brain-inspired, foundational innovations to the software and hardware architecture of future fault-tolerant and dynamic many-tile systems, to be applied to those Embedded Systems and High Performance Computers requiring extreme numerical and DSP computation capabilities.

The project will deliver:
- Experimental Many-Tile Platforms for the study of many-tile systems in the scenarios of Embedded Systems and HPC (QUonG Hardware Platform and VEP-EX simulation platform);
- A many-tile programming/optimization environment, to be applied to dynamic, fault-tolerant, many-process numerical/DSP applications, with foundational innovations;
- A set of application benchmarks, representative of both HPC and Embedded System domains, coded using the new programming environment.

### 3.1.1. Software Tool-Chain

From a software perspective, EURETILE aims at providing a scalable and efficient extensive programming framework for many-tile platforms, by exploiting the underlying parallelism and investigating some key brain-inspired architectural enhancements specific to EURETILE. The proposed programming environment is based on a new model of computation that explicitly exposes the coarse-grain and fine-grain parallelism present in applications. Then, automatic tools are aiming at obtaining predictable and efficient system implementations by optimally matching the concurrency and parallelism present in applications, with the underlying many-tile hardware. The figure sketches the envisioned software EURETILE tool-chain. The software tool-chain follows the well-known Y-chart approach and in EURETILE each design phase will be enriched with novel concepts for programming efficiently the three layers of hierarchy present in EURETILE platform. The proposed software tool-chain aims at optimizing issues related to system throughput while still guaranteeing real-time constraints for applications that operate under such restrictions. Additionally, fault-tolerance aspects are considered, from system-level programming. The way of representing parallelism, concurrency, and fault tolerance at system level, the implementation of the distributed real-time operating system, and the efficient fast simulation environment have all a strong impact on the overall optimality of the system implementation.

The proposed programming framework developed by ETHZ is based on a new model of computation that matches well the many-tile EURETILE architecture. We have identified three levels of hierarchy in the "network of processes" describing the structure of the application, in analogy to three levels identified in the cortical-inspired organization of the hardware, i.e., 1-cortical columns, 2-cortical areas, and 3-neo cortex. The two upper levels of hierarchy in the network of processes are used to represent the coarse-grain concurrency. The first level is associated to finer grain parallelism. Practically, the concept of a distributed operation layer (DOL), applied with success to the multi-tile SHAPES platform (see SHAPES FP6 project http://apegate.roma1.infn.it/SHAPES and http://www.tik.ee.ethz.ch/~shapes) will be extended to describe the tremendous concurrency of different sets of parallel applications running simultaneously (dependently or independently) on the new brain-inspired many-tile architecture. In SHAPES, DOL was applied to describe static stream-oriented applications and for mapping only a single application to a multi-core architecture. In other words, the third level of hierarchy as described above was missing. Moreover, in EURETILE we aim at executing concurrently several dynamic applications, taking advantage of the many-tile distributed platform.

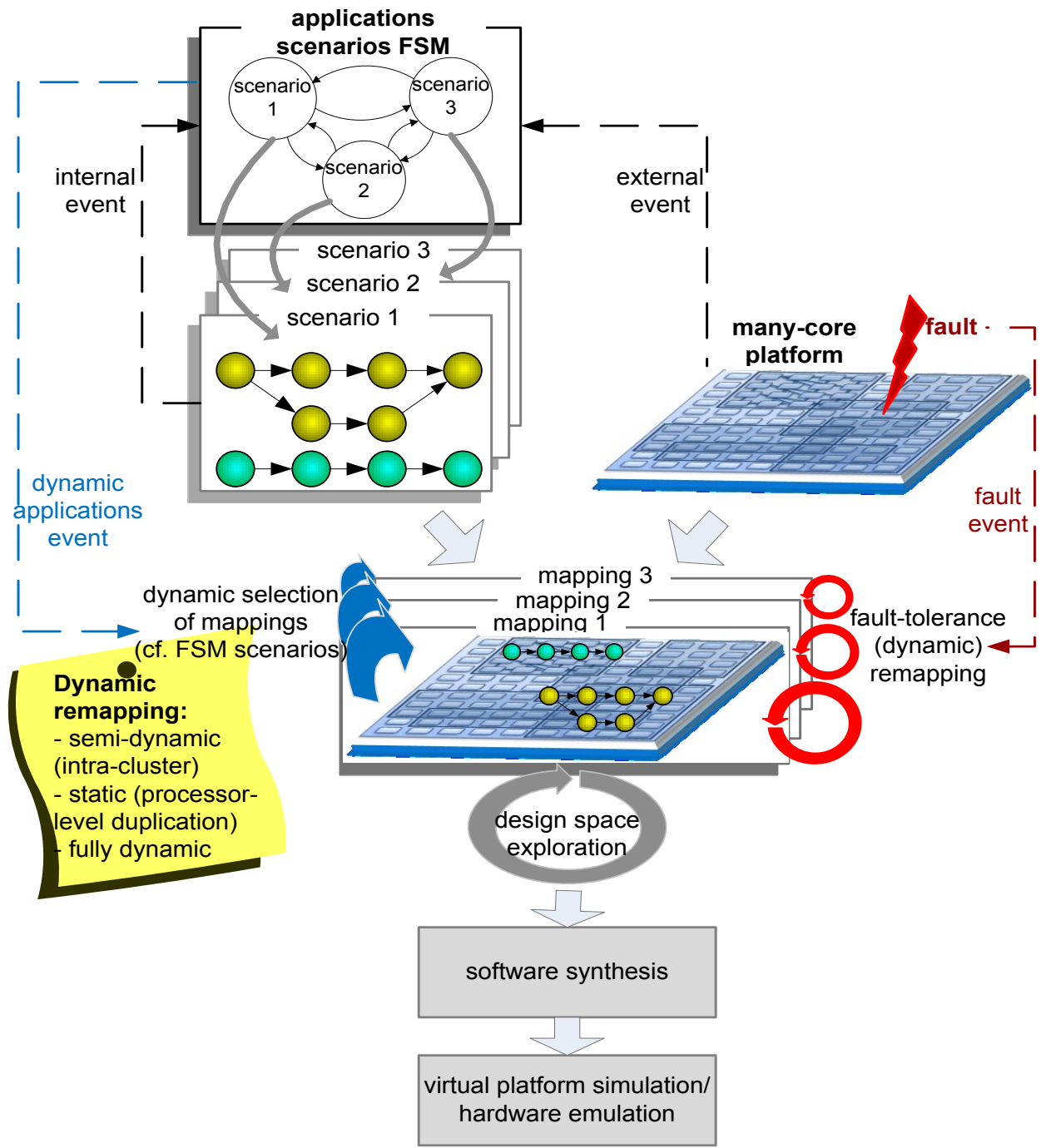

**Figure 3-1. Schematic View of the Holistic and Scalable EURETILE Software Tool-Chain.**

To answer all these challenges, the DOL programming environment needs to be enriched with innovative concepts. First, a new programming model that we will call Distributed Application Layer (DAL) is being developed at ETHZ, corresponding to the third level of hierarchy of EURETILE platform.

This adds one level of hierarchy above the current process network layer in DOL, allowing the application programmer to specify concurrent execution of several dynamically instantiated applications on the same hardware.

Second, the DOL programming model, which still relates to the first two levels of hierarchy in the architecture, needs to match well the underlying brain-inspired hardware platform. In particular, planned and unplanned internal and external events will be considered as initiators of dynamic remapping, exploiting the hierarchical nature of the underlying execution platform and the dynamism and concurrency inherent to the applications. (Semi-) automated off-line and efficient on-line mapping methods are to be investigated, including run-time monitoring and optimizations that consider timing predictability of individual applications as well as efficiency in terms of system throughput and accelerated executions.

Then, in addition, fault tolerance needs to be part of all aspects of the EURETILE system, i.e., hardware platform, programming model, compilation, hardware-dependent software, operating system, and many-tile simulation. Fault-tolerance aspects are concretely taken into consideration starting from system-level, in the EURETILE programming environment and dynamic mapping strategies on many-tiles. In this sense, the DAL environment allows designers expressing fault-tolerance aware mechanisms by dedicated programmable routines.

### 3.1.2. Experimental Hardware and Simulation Platforms for Embedded Systems and HPC

The project develops two platforms, which will be used to experiment innovations to future many-tile software and hardware architectures:

- VEP-EX, Embedded Systems many-tile simulation platform. The simulation framework will be developed by RWTH-AACHEN and will include multiple RISCs networked through a custom interconnect mesh composed of INFN DNPs (Distributed Network Processors).
- The QUonG Hardware Platform for scientific high performance computing. The hardware platform is networked through a custom interconnect mesh composed of DNPs hosted on FPGA boards (APENet+, providing innovative fault-awareness and fault-injection features). The platform includes off-the-shelf boards mounting INTEL multi-core CPUs. INFN will also explore the addition of GPGPUs.

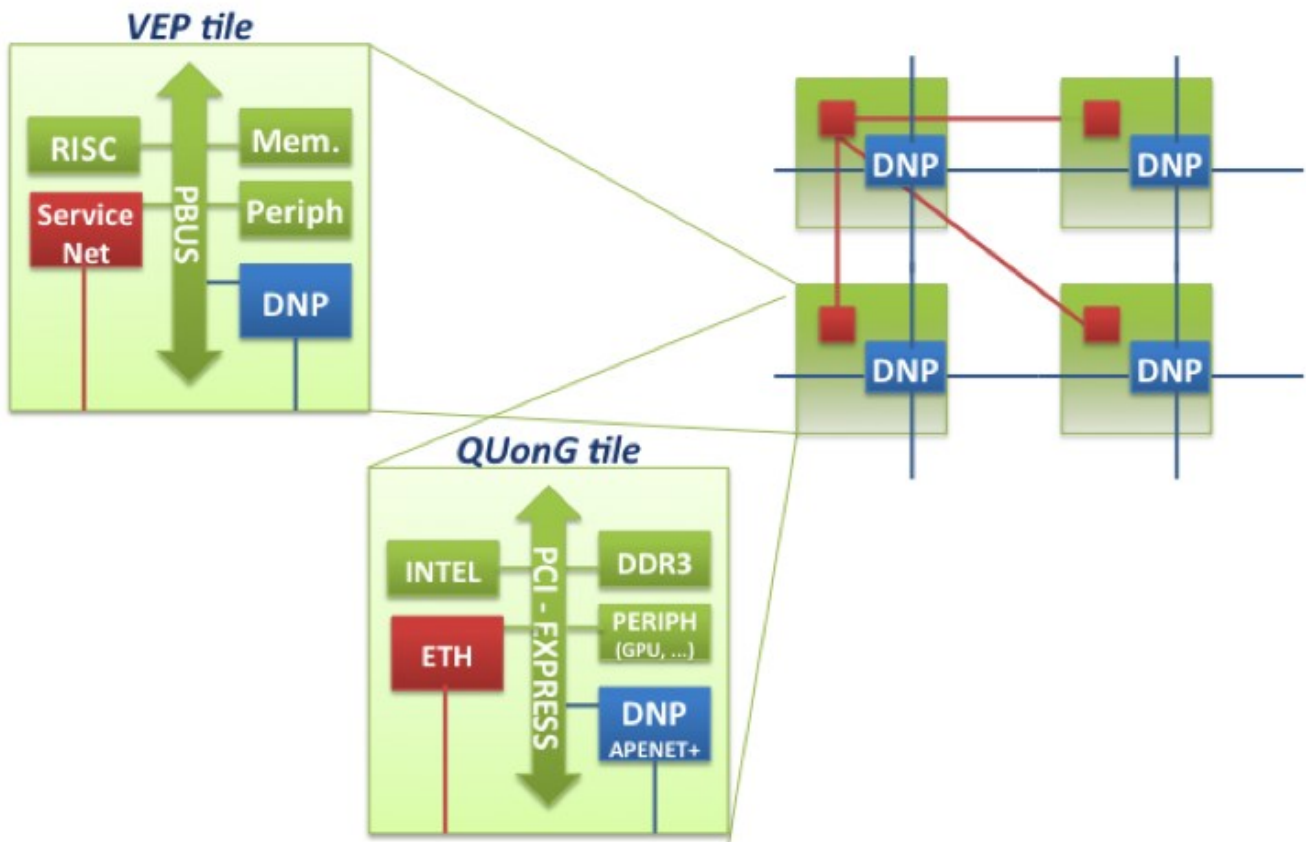

**Figure 3-2. Simplified view of the EURETILE platforms (VEP and QUonG), which will be used for experiments on hardware and software innovations on future many-tile Embedded Systems (VEP) and HPC (QUonG).**

The project will also explore the addition of software programmable hardware accelerators in the form of ASIPs, using the framework developed by TARGET Compiler Technologies.

### 3.1.3. Definition of a Hierarchical Brain-Inspired Many-Tile Many-Process Architecture

The plan is to match the coarse-grain and fine-grain concurrency of applications with that of the underlying hardware platform, to obtain predictable and efficient systems.

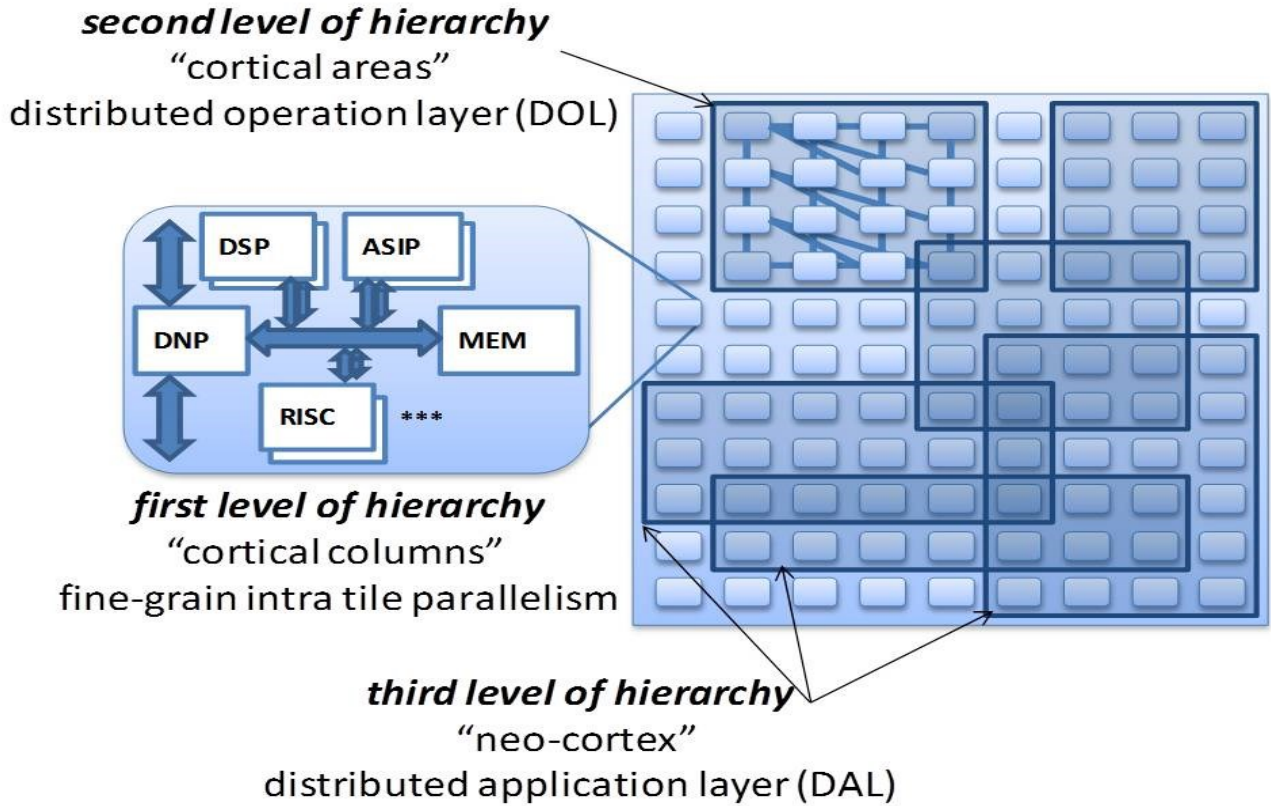

**Figure 3-3 Hierarchical Brain-Inspired Architecture: Conceptual View, applied to the Embedded System case. At the first (bottom) level of hierarchy, the elementary tile includes the DNP (Distributed Network Processor) for inter-tile communications, general-purpose processor(s) and local memories. Depending on the target application domains, the tile can include DSP(s) and ASIP(s). Upper levels of the hierarchy are constructed according to the distributed memory paradigm.**

In particular, we will research on reflecting the three levels of hierarchy of the brain-inspired architecture (1-cortical columns, 2-cortical areas, 3-neo cortex), in the dynamic network of processes that describes the structure of the application.

- **First level of the hierarchy: "Cortical columns" vs. EURETILE elementary tile.** The computational kernel carried on by each process can access in read/write mode all the variables stored in the tile memories, in analogy with the "columnar computation" enabled by the all-to-all network among the neurons inside the column. The fine-grain concurrency (instruction-level parallelism and shared-memory threads) will be exploited at the level of abstraction corresponding to complex numerical computations, elementary signal or image processing operators, filters, small scale FFT or matrix computations, inversions, state machines or any other single-thread of computation, including the simulation of a set of neighbouring neurons.
- **Second level of the hierarchy: "Cortical areas" vs. EURETILE arrays of hardware tiles and processes.** In EURETILE computations, it will be possible to declare arbitrary small and large-scale communication patterns between processes (such as n-dimensional arrays of message-passing connected computational processes), which will then be mapped on arrays of computational hardware tiles. This kind of n-dimensional decomposition of data-structures into smaller domains is a classical strategy of DSP and numerical algorithms.

- **Third level of the hierarchy: "Neo-cortex" vs. EURETILE hardware system and complex software application description.** On EURETILE, it will be possible to program a broad range of complex numerical, DSP and control applications coding them as a network of high abstraction computational models, as described in the second level of hierarchy. Several of these mostly independent, but possibly communicating applications, will be concurrently executed and share the hardware platform. This layer is dynamic in the sense that applications may dynamically enter or exit the system.

The figure graphically represents the different levels of brain-inspired hierarchy and the associated concepts in concurrency management, applied to the Embedded System case.

### 3.1.4. Fault-Tolerance and Scalability

Elementary processors are designed to guarantee a failure rate low enough to sustain the execution of platforms based on few components. The scaling to many processor systems requires the management of more frequent failures. Current parallel systems manage faults with a primitive approach based on period check-pointing (i.e. periodically saving the full state of the application on external storage) managed by the application programmer. If a hardware fault happens, the hardware is repaired or excluded, and the previous checkpoint is reloaded, as soon as the hardware system or a partition is available. In EURETILE, the aim is to automate the handling of the faults. From hardware point-of-view, the DNP (Distributed Network Processor) will be able to discover critical events, faulty lines and/or faulty tiles, and to inject synthetic critical and fault events, stimulating a systemic fault-management reaction. The DNP will monitor the system behaviour, and will signal the failure to the software environment, for automation of higher-level strategies. We will also investigate the routing of messages around faulty components.

A key topic is the investigation of the scaling behaviour of the brain inspired system to many-tile configurations. To such purpose it will be useful to raise the level of abstraction and use a simulator capable to show the behaviour of high abstraction models of DNPs and processors under different kind of synthetic traffic workloads and failures.

### 3.1.5. Versatile Distributed Network Processor

The DNP will be compatible with the platforms above described through appropriate processor interfaces. A more flexible software-programmable DNP can be designed in the form of an ASIP (using TARGET's ASIP design tool-suite). The hardware prototype delivered by EURETILE will be integrated using a custom interconnection integrated on FPGAs. In principle, DAL (ETHZ) will be compatible with all proposed platforms, due to the platform-independent characteristics of this system-level environment. However, in order to apply DAL to effectively programming EURETILE platforms, fundamental platform-specific support need to be implemented in the hardware-dependent software (UJF-TIMA), and ensued by simulation capabilities of the fast simulation environment.

### 3.1.6. ASIP Design

The HW/SW co-design tools operate at the level of elementary tiles in the brain-inspired many-tile architecture. Its work is based on its retargetable tools-suite for the design and programming of Application-Specific Processors (ASIPs). Building on this technology, Target contributes to the project in two major ways. First, new software-programmable accelerators will become part of the elementary tiles. These accelerators take the form of ASIPs optimised for the typical numerical kernels of the applications envisaged in the project. Architectural exploration is based on profiling of application code in the retargetable SDK. In addition, the tools generate efficient RTL hardware models of the ASIP, enabling a quick implementation on FPGAs. Secondly, Software Development Kits (SDKs) will be developed for any processors in the elementary tiles that are intended for numerical data processing, which may include existing DSPs or new ASIP accelerators. Key elements of these SDKs are an efficient C compiler and an on-chip debugger. Third, ASIP will contribute to the acceleration of the DNP/APENet+ interconnection system.

### 3.1.7. Application Benchmarks

A few characteristics of the project determined the selection of the set of benchmarks:

- The software environment and the hardware architecture conceived by EURETILE should be qualitatively and quantitatively driven by the solution of problems selected from both the embedded systems and scientific computing scenarios;
- EURETILE is about future many-tile architectures applied to dynamic scenarios that include numerically intensive digital signal processing and scientific kernels, individually represented by Kahn many-processes networks;
- We should explore dynamism and fault-tolerance of many-processes applications;
- We also expected to exploit hints coming from the brain architecture, a system capable of extreme parallelism and low power operation.

We decided to fix the following set of benchmarks, selected from 3 application domains:

1. Dataflow oriented / digital signal processing benchmarks. The new DAL environment will be used to describe the expected dynamism (changing scenarios, e.g. for mobile appliances), using an explicitly tiled description of the available parallelism. This set of benchmarks is developed by ETHZ, a partner with a consolidated experience in the solution of optimization techniques for embedded systems.
2. The DPSNN-STDP (Distributed Polychronous Spiking Neural Net with synaptic Spiking Time Dependent Plasticity). cortical simulation benchmark, used with 3 purposes:
   a. as a source of requirements and architectural inspiration towards extreme parallelism;
   b. as a parallel/distributed coding challenge;
   c. as a scientific grand challenge.
   The study of the brain simulation benchmark will be developed under the direct responsibility of the coordinator (Pier Stanislao Paolucci – INFN Roma).
3. LQCD (Lattice Quantum Chromo Dynamics), a classic HPC grand challenge, which since 1983 had been driving the development of several generations of massive parallel/distributed computers. This benchmark will be maintained by the APE group (INFN Roma).

The Brain Benchmark will be coded during the temporal framework of the EURETILE project both using a standard C/C++ plus MPI environment, as well as using the C/C++ plus XML DAL environment. This will allow to compare the features of the new environment in comparison to a classical environment on a benchmark which will be coded "from scratch" using an explicit description of parallelism.

### 3.1.8. Main expected final results

To achieve this goal, our main **expected final results** are:

- The definition of a hierarchical, scalable many-tile HW architecture, taking inspiration from the cortical structure, and searching for brain-inspired and fault-tolerant foundational innovations;
- A hardware prototype (QUonG) integrating at least 64 tiles and 128 cores, with support for fault awareness and tolerance;
- A scalable and flexible simulator, enabling the exploration of behaviour under faults and an easier debugging of many-process applications;
- A set of highly representative benchmarks: LQCD (Lattice Quantum Chromo Dynamics), PSNN (Polychronous Spiking Neural Networks), multi-dimensional FFT (Fast Fourier Transforms) and DSP/Data Flow Oriented, demonstrating the potential of EURETILE architecture on a broad class of numerical, DSP, and control applications on both scenarios (Embedded Systems and HPC);
- A many-tile programming environment, where applications can be expressed as dynamic network of processes and can be dynamically and mapped and controlled in an efficient manner on the 3-levels brain-inspired architecture (cortical columns, cortical areas, neo-cortex);
- The implementation of many-tile awareness of critical events and faults, and fault-tolerance at system-level through the support of all software and architecture layers.

### 3.1.9. Work-packages

The project is structured in ten work-packages:

| WP1 | Brain-Inspired Many-Process System Software Requirements |
| WP2 | Many-Tile Hardware Architecture Specification |
| WP3 | Foundational Many-Process Programming Environment |
| WP4 | Distributed Hardware Dependent Software Generation |
| WP5 | Many-Tile Simulation/Profiling |
| WP6 | Innovations on Hardware Intellectual Properties |
| WP7 | Challenging Tiled Applications |
| WP8 | Software Tool-Chain Integration |
| WP9 | Training, Exploitation and Dissemination |
| WP10 | Management |

## 3.2. **Work Performed and Achieved Results in 2010 (First Year)**

*This section is a record of activities and results obtained during the first year of the project, and the technical content has been maintained as in the original yearly reports to maintain an historical track of the project flow. Next sections (3.3. and 3.4) describe the project evolution in 2011 and 2012.*

### 3.2.1.  HW Architecture Design

During the first year the HW Architecture design has been performed by INFN in cooperation with the consortium partners.

WP2 led by INFN and including contributions from other partners, started and concluded during this first period and results are reported in the deliverable D2.1.

The overall objective of WP2 was to produce a preliminary system specification of the EURETILE many-tile hardware platforms. The collaboration decided to go for 2 parallel and synergic development lines: the first in the area of embedded systems while the other in the area of HPC systems.

From the software point of view, the unifying elements are the common software tool-chain (from the high level programming mechanism to the HDS) and the simulation framework while the main hardware related developments is in the area of scalable, fault tolerant, brain-inspired network capable to be interfaced to custom ASIP, optimized for specific scientific applications (specifically for LQCD and PSSN in the EURETILE case studies), and/or commodities computing accelerators.

### 3.2.2.  DNP Design

This area of activity is covered by INFN. In particular, in the area of DNP enhancement main goals for 2010 were:

- Design of a DNP-based 3D Torus network for HPC systems through:
  - VHDL Coding of DNP interface to HPC PCI Express based host system
  - DNP-uP integration to optimize RDMA support for INTEL-based host system
- Porting of DNP 3D Torus network controller on FPGA platform
- Design of a 6 channel FPGA-based electronic card for HPC system

During 2010, DNP activities were executed in collaboration with the APEnet+ project. The INFN initiative called APEnet aims to build high performances custom networks for HPC commodities PC Clusters using ideas, know-how and IPs developed in the framework of the APE project.

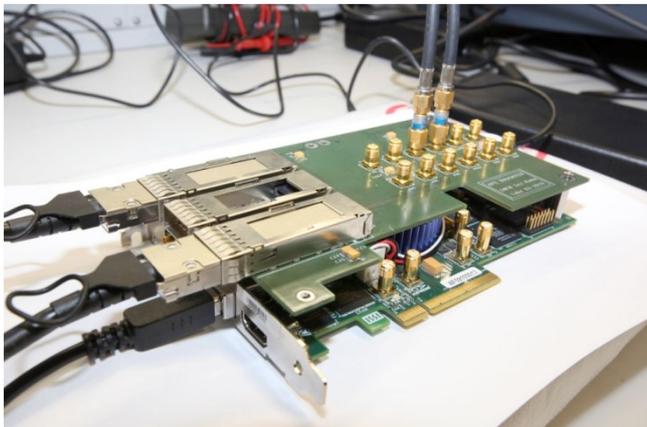

APEnet+ is the last generation of NIC boards (FPGA based) implementing a 3D Torus network for PC Cluster.

During 2010, we progressed toward the EURETILE HPC platform, which foresees the development of an FPGA-based network card (a 3D Torus network router using the DNP architecture) to interconnect commodities PC clusters. During this first year, we synthesized the current DNP on the target FPGA ALTERA STRATIX IV EP4SGX 290 device obtaining preliminary but encouraging results (clock frequency and footprint). Then, we redesigned the DNP serial-link interface to be compliant with the target Ser/Des, for a total speed of 34Gbps per toroidal link. The DNP-PCI Express interface was developed and a preliminary test of integration of microprocessor for better support RDMA operations was performed. A complete design of the APEnet+ prototype board was the result.

**Figure 3-4. DNP test board (APEnet+) developed during 2010.**

### 3.2.3. LQCD and PSNN ASIP Design

During the first year of activity, the ASIP SDK technology has been extended by TARGET with new methods for checking the suitability of a processor model for C compiler support, advanced on-chip debugging capabilities and a new graphical integrated development environment. Then, an initial feasibility study was made on the design of two ASIPs for the scientific numerical processing tasks envisaged in the project (LQCD and PSNN).

### 3.2.4. System-Level Programming Framework

In 2010, DAL and its main concepts have been specified and the corresponding DAL API has been established. The DAL specification document, including the basic DAL description, API, and interfaces with other work-packages has been sent to all partners, and the key issues are summarized in deliverable D1.1 "Report on Brain-Inspired Many-Process System SW Requirements" (delivered at M12). During 2010, we developed a first implementation in a distributed functional SystemC simulation that can check the correct functionality of multiple concurrent applications, dynamic mapping and re-mapping strategies, as well as some fault-tolerance features. For the following EURETILE project periods, (i.e. starting from 2011) we planned an advanced implementation of the distributed functional simulation to be used by all partners, as basis for a bug-free application development, support for the final HdS implementation, and source of first results in terms of performance analysis and system optimization.

### 3.2.5. Scalable Simulation Environment

This area of activity is led by RWTH, in cooperation with the other partners. The main target of the scalable simulation environment is to support the development and validation of system hardware/software and applications. This activity comprises four different areas: Abstract simulation, parallel simulation, debugging support and fault injection. Since the beginning of the project, work on the following topics has been performed:

**Abstract simulation:** The Hybrid Processor Simulation (HySim) has initially been deployed on ARM and magic processors within the SHAPES project. In the meantime, the HySim engine has been restructured for easier porting and improved to support more complex applications, such as those that will be part of EURETILE. In addition, an extension to this concept for the purpose of fast Network-on-Chip simulation has been investigated on an exemplary tiled platform. A satisfactory initial model has been tested and a full integration in the simulation flow will follow.

**Parallel simulation:** In 2010, research on parallel simulation that was started during the FP7 SHAPES project has been continued. Furthermore, a mixed-level parallelized simulation of a tiled computing platform has been created. It consists of basic tiles with processing elements and local memory, which are connected by packet routers. The simulation was exhibited at DAC 2010, and the technical details were presented at CODES/ISSS 2010.

**Debugging support:** Novel debugging techniques to lower the complexity of MPSoC software development on the EURETILE system were specified and presented to the project partners. Such techniques are based on the advanced visibility, controllability and determinism of virtual prototyping, and aim at finding software malfunctions caused by concurrent inter-tile and intra-tile activity. The implementation of debugging support to cope with current simulation technology used in EURETILE will be covered during next project stages.

**Fault injection:** Discussions are on going with hardware and system software partners, regarding which kinds of faults can be handled by the EURETILE system. These faults will then be modelled and supported for injection using the system simulator implemented.

### 3.2.6. Hardware-Dependent Software

UJF-TIMA is responsible of providing HdS for the SW tool-chain able to generate binary codes for processors on the HW platform. We define hardware-dependent software (HdS) to be "software directly dependent on the underlying hardware." UJF-TIMA is in charge of providing HdS, which includes OS services and low-level software.

The main challenges for HdS generation come from the brain-inspired paradigm and the hierarchical architecture. This may imply some dynamic changes during execution, such as new task running on processor for new application or a failure detected somewhere in the architecture, as the brain is naturally fault-tolerant with redundancy. For that reason, we are dealing with fault tolerance aware software services provided by HdS and real time properties as well as new control mechanisms. New control mechanisms concern the overall architecture and control organization: distributed, centralized or a hybrid version. In 2010, this has been still an on-going work.

An emerging solution for fault tolerance seems tasks migration. But this is really challenging in NUMA architecture, non-SMP, and in a hierarchical architecture. Task to be migrated is supposed to have reached a specific point of execution, and a copy of all what is needed to restart this task on another processor (stack, registers, memory, task code, cache?) will be sent. The starting point of EURETILE is an in-house OS (called DNA-OS developed for the previous FP6 SHAPES project) running on each processor of the architecture. During 2010, DNA-OS has been extended to include all what was missing for SMP system and task migration, that was the first step to move forward the EURETILE context. DNA-OS is running on ARM processors, as well as other architectures (SPARC V8, MIPS R3000, mAgicV,)

Control mechanisms and task migration are based on an efficient communication infrastructure and SW drivers. This has been consolidated with an automatic driver selection and specialization regarding communication protocol selected and requirements.

## 3.3. **Work Performed and Achieved Results in 2011 (Second Year)**

*This section is a record of activities and results obtained during the second year of the project, and the technical content has been maintained as in the original yearly reports to maintain an historical track of the project flow. Next sections (3.4) describe the project evolution in 2012.*

### 3.3.1. **Definition of Fault Aware Architecture**

During 2011, INFN proposed to the consortium a novel HW design paradigm, named "LO|FA|MO", which creates a *systemic awareness of faults and critical events*, thanks to a distributed approach that uses additional hardware components on each DNP, and needs dedicated software components running on each tile to create a *local awareness of faults and critical events*. This local awareness is then propagated along the system hierarchy. Onto such local and systemic fault awareness, a software approach to fault reactivity can be grounded, which has been proposed by ETHZ at the end of 2011. From a system-level perspective, two fault reactivity strategies have been investigated in WP3 and first ideas have been implemented in the Distributed Application Layer (DAL), namely *fault recovery* and *fault tolerance*.

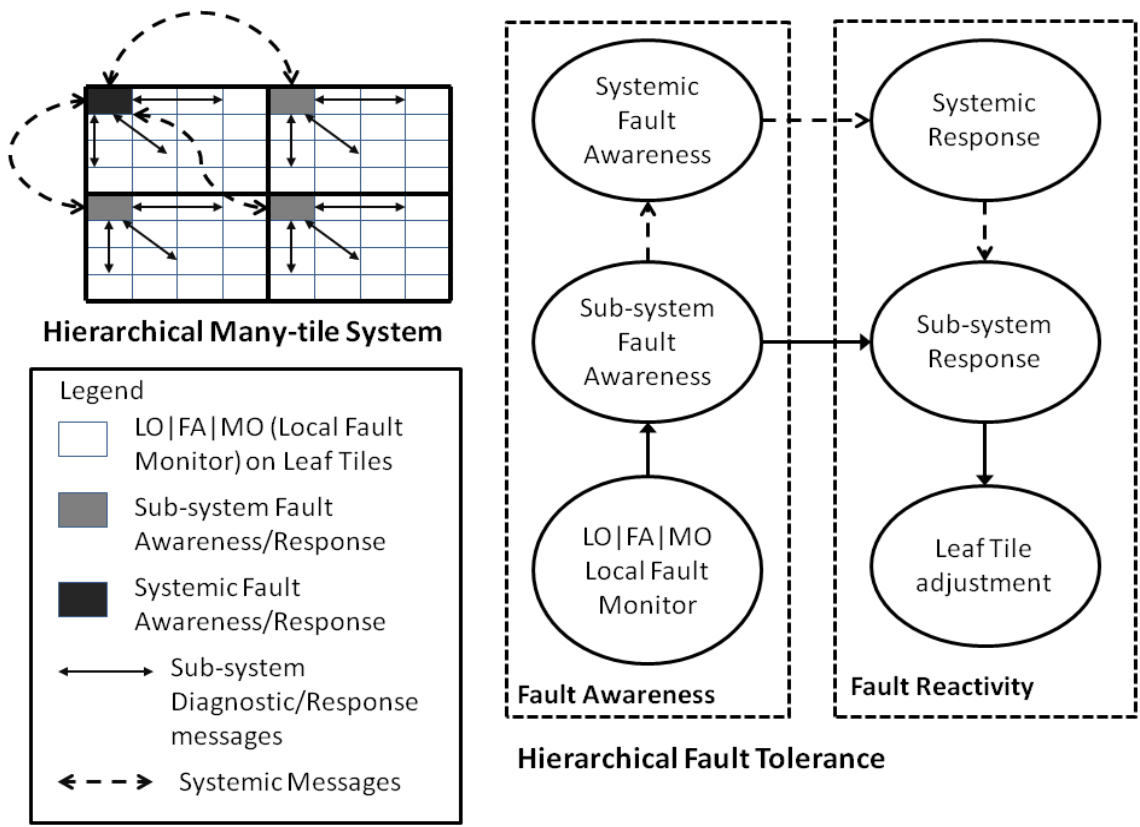

**Figure 3-5 - The Network Processor of each leaf in the many-tile HW system is equipped with its own LO|FA|MO components. The *Local Awareness* of faults and critical events is propagated towards the upper hierarchy levels, creating *Systemic Awareness*. Reactions to faults and critical events could be autonomously initiated by sub-system controllers or could require a systemic response.**

### 3.3.2. **Challenging Application Benchmarks**

We have three benchmarking areas: DSP/Data Flow Oriented, PSNN-Brain Simulation and Lattice Quantum Chromo Dynamics.

During 2011, several application benchmarks have been implemented as distributed application layer (DAL) specifications.

About **DSP/Data Flow Oriented**, at ETH Zurich, several DAL-compliant applications have been specified, of which the most significant are a picture-in-picture application consisting of two parallel MJPEG decoders, and the kernels of a fast-Fourier transformation, and matrix multiplication. These applications are included in the DAL distribution and can be downloaded from http://www.tik.ee.ethz.ch/~euretile.

Topologically, the applications consist of a task-level parallel specification in the form of individual process networks and an application-level parallel specification, expressing dynamic application scenarios, in the form of a finite state machine. The DAL framework is then capable of deciding on the mapping or remapping of applications/processes as well as to investigate some key features related to application's fault management. In particular, case studies of the picture-in-picture application have been designed to demonstrate the parallel execution of multiple applications, as well as basic DAL actions such as starting, stopping, pausing, and resuming the different videos, together with the implementation of the control infrastructure. Moreover, distributed mapping and remapping of individual component processes are investigated. Basic fault management can be illustrated at mapping level, where applications residing on damaged cores are smoothly redirected towards reserved healthy cores.

These investigations open the way for further research and design options, such as process level redundancy for critical applications, and the on-line exploitation of pre-calculated performance trade-offs.

**PSNN**. About the study of Brain Simulation, used as benchmark and as a drive source of architectural concepts, during 2011 we produced a first version of D-PSNN, the natively Distributed – Polychronous Spiking Neural Network Benchmark. The benchmark has been compiled and run to initialization a "toy-scale" network in two frameworks: the DAL environment (C++ plus XML) and a "standard" C++ plus MPI context. The coding has been executed after:

1- The production of a list of "computationally Inspiring" review of biological facts from experimental and theoretical neuroscience;
2- A complete "reverse engineering" of a sequential reference PSNN code, to grasp all details and understand the blocking point against parallelization.

**LQCD**. The INFN team has a multi-decennial experience on Lattice Quantum Chromo Dynamics, and a multi-process representation of the problem available. Starting from 2012, LQCD will be used to validate the new features associated to dynamism and fault tolerance developed by ETHZ in 2011. During this year, we performed a quantitative reassessment of bandwidth and latency requirements, taking in account the characteristics of the EURETILE HW platform.

### 3.3.3. Innovation on HW Intellectual Properties and design tools

In 2011, in addition to the definition and first implementation of the LO|FA|MO fault awareness system, INFN main achievements were the delivery of a refined DNP component and the introduction of APENet+ and QUonG hardware prototypes. On the ASIP development side TARGET designed a specific ASIP architectures, optimised for application domains targeted by the project.

The DNP released at the end of 2011 is a powerful, robust and faster design thanks to the introduction of specialized hardware to speed-up PCIe side transactions as well as the 3DTorus transfers.

The APENet+ final board, integrating the last release of DNP IP, was produced in pre-series (4 boards) and validated with extensive test sessions.

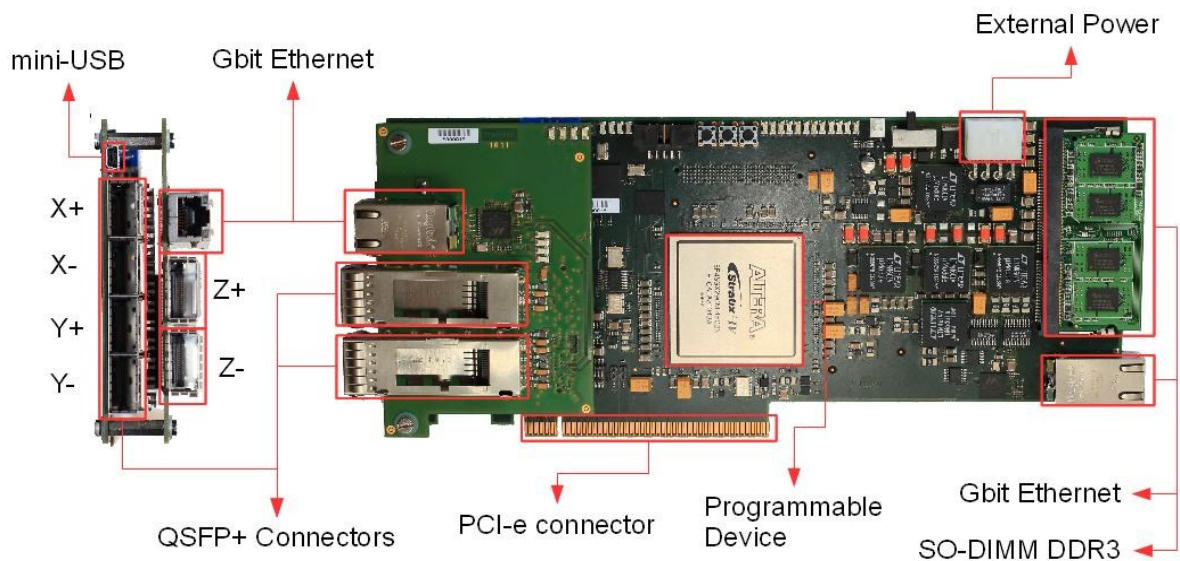

**Figure 3-6. APENet+ Board produced in 2011.**

INFN also put in production a reduced prototype of QUonG system, the hybrid parallel cluster selected as the HPC Euretile platform.

### 3.3.3.1. DNP Innovation.

During 2011 INFN executed several refinements of the DNP IP in order to get a more robust and faster design.

In particular we focused on the optimization and improvement of the "network interface" block adding dedicated hardware logic to improve RDMA tasks execution and speeding-up the recently introduced hardware block implementing GPU *peer-to-peer* protocol (a very brand new concept for the current commercial network cards).

Some work was also devoted to the enhancement of physical links interface and to the optimization of the FPGA embedded microprocessor (NIOSII) firmware.

In addition we also started to define the architecture and planning the design of the hardware blocks required to implement the LO-FAMO fault tolerant mechanism in the EURETILE platform demonstrator.

**APEnet+ innovation**. INFN introduced the first release of the APEnet+ board, the DNP-based, PCIe board implementation of the 3D Torus network, featuring 6 fully bidirectional links with 34 Gbps of raw bandwidth per direction. With the integration of the P2P Enabled DNP into its high-end FPGA, APEnet+ is the first non-NVidia device with specialized hardware blocks to support the NVidia GPU direct peer-to-peer inter-GPU protocol. During the 2011 we produced and tested a set of 4 APEnet+ prototypes on which we performed a preliminary bandwidth and latency analysis obtaining very good results in terms of access latency reduction.

**QUonG innovation.** QUonG, the successor of the former APE supercomputer and the prototype of the HPC EURETILE platform, is a hybrid INTEL server x86-based cluster accelerated using NVidia GPUs and interconnected by APEnet+ 3D Torus network.

In 2H11 we initiated the procurement phase of QUonG reduced prototype (25 TFlops) to be upgraded to a full "QUonG tower" system, capable of a peak performance of ~65 TFlops, during 2012 and 2013. The QUonG tower will be made available to the Euretile collaboration to perform tests and software development.

### 3.3.3.2. ASIP Design Tool Improvement.

The Euretile consortium's approach towards the design of ASIP is to use a retargetable tool-suite based on a processor description language. The tool-suite supports architectural exploration, the generation of a software development kit, and the generation of an RTL hardware implementation of the ASIP. TARGET's IP Designer tool-suite is used for this purpose.
In project year 2011, TARGET has developed a number of new extensions and optimizations for its retargetable tool-suite. These activities are part of WP6. These developments focussed on two main topics:

- A new methodology has been defined to model memory and communication interfaces of ASIPs.
- New techniques have been developed and implemented to provide enhanced feedback to users of the retargetable tool-suite during the architectural optimisation process.

### 3.3.3.3. LQCD ASIP Design.

In project year 2011, we started the development of specific ASIP architectures, optimised for application domains targeted by the project. These activities are part of WP6, and were carried out by TARGET in cooperation with INFN. More specifically, an efficient ASIP for LQCD has been designed using the IP Designer tool-suite. The ASIP has been named "VCFLX" (referring to Vector Complex Floating-point − Flexible"). Benchmarking shows that a single VCFLX ASIP can implement LQCD 33x faster than a 32-bit floating-point CPU and 4x faster than the mAgicV DSP. About 50 instances of the VCFLX ASIP can fit on a single Virtex-7 FPGA, thus speeding up LQCD even further, provided that the FPGA's RapidIO fast communication channels can be used to achieve a sufficiently high communication bandwidth.

### 3.3.4. System-Level Programming Framework (DAL)

During 2011, in the frame of WP3 led by ETHZ, the second complete prototype of the distributed application layer (DAL) has been implemented and the consolidation of DAL underlying concepts has been done in line with all project work-packages and tools.
Distributed application layer (DAL) is a system-level programming environment for many-core architectures that considers dynamic application scenarios, i.e. applications that may appear and disappear dynamically at runtime. At the same time, the approach allows designers to handle at system-level faults that occur at runtime. In particular, besides mapping optimization of applications to the EURETILE platform, DAL provides the environment for remapping or restarting applications in case of changes in the application scenarios or in case of permanent system faults. However, providing timing guarantees for applications remapped/restarted due to faults that occur arbitrarily at runtime is hard. Therefore, in DAL additional methods are investigated such that specification, generation, and analysis of redundant task graphs for so-called critical applications with real-time constraints.
As a result of 2011 efforts, the second DAL prototype has been made available including the complete specification, a functional simulation that is automatically generated from the initial specification, several case studies demonstrating the capabilities of DAL, and a complete documentation, see http://www.tik.ee.ethz.ch/~euretile. As well, to make the tool-chain complete, basic tools are included for performance analysis, search in the design space, and mapping optimization under specific platform constraints.

The key challenges encountered by application programmers when designing applications using DAL are in a fist instance exposing the process-level parallelism in the application, and second, exposing application-level concurrency and dynamism in a formal programming model. Additional challenges are in exposing fault-awareness of the system early in the design and deciding among proposed fault management strategies. To answer all these challenges, the DAL functional simulator has been developed as a tool for application programmers, to test and profile their applications early in the design, on top of an ordinary computer system.

Technically, the DAL functional simulator version made available at the end of 2011 has been running on top of Linux and uses POSIX threads to execute DAL processes that simply communicate via the shared memory of the Linux system. Additionally, a specific process network designed for the hierarchical control of application-level dynamism is first computed out of the specification, automatically generated during the synthesis step, together with the functional simulator, and then loaded when the system is booting-up. The design space exploration tool provides the mapping information. It consists of a set of mappings, each one valid for a set of application states. This set is compiled into the master controller that provides the basic process management instructions for the slave controller, such as starting, pausing or restarting a task as well as installing communication channels and the corresponding memory.

In the proposed functional simulator virtually all main DAL concepts can easily be tested. Therefore, such a functional simulator is enabling the fast development and test before porting the DAL design to the target EURETILE platform. Moreover, such a fast development prototype is opening the way for further research in the areas of performance analysis, mapping/remapping optimization, and fault-awareness offering all functional capabilities of the EURETILE system, and hiding all low-level details and parallel development of the hardware-dependant software.

### 3.3.5. Scalable Embedded Computing Simulation Environment (VEP)

During 2011, work has been performed by RWTH related to the simulation environment on the following topics:

**VEP-EX simulator:** The base version of the VEP-EX simulator has been realized and distributed to the partners. The VEP-EX simulator offers two different selectable computing elements: Either the IRISC processor instruction set simulator is used, or alternatively an abstract execution device (AED) is employed.

The AED is a SystemC module that can load host-compiled shared libraries containing code to execute inside the simulator environment. This mechanism is used to ramp up DNP drivers and test software before the actual tool-chain for the RISC processor becomes available. Introducing the AED is additional effort not originally part of the project plan. It has become necessary as a bridging solution until the RISC software tool-chain becomes available.

**Debugging support:** A debug technique for concurrent software running on heterogeneous multi-core systems was developed in order to set the grounds for a preliminary debugger framework for the EURETILE system. This new technique, named Event-based Bug Pattern Descriptions (EBPD), adds automation for debugging by using assertional techniques similar to those in formal verification and can be easily retargeted to cover different programming models and systems. Some building blocks of the underlying debugger architecture, like the BPD grammar, the BPD compiler, the Event Monitor (EV) and the Sequential Consistency Analyser (SCA), were developed and evaluated in simpler multi-core systems. Common bug patterns in concurrent software were also described using the BPD grammar in order to start populating a database of well-known patterns.

**Abstract simulation:** The HySim technology, legacy of SHAPES, has been enhanced in two ways. Firstly, the framework has been redesigned to support a class of conceptually different processor architectures, namely customizable cores, and a more profound way of processor state synchronization between the native and the target processors has been introduced. Secondly, a time synchronization mechanism has been introduced, to enable the support for multi-core or any platforms relying on timing, like those including watchdogs or timeouts. Apart from that, some preliminary experiments were conducted on a multi-core TLM2.0 platform, to evaluate the hybrid NoC simulation.

 **Parallel simulation:** In collaboration with Synopsys, Inc., parallel simulation has been further researched, with focus on applicability to systems currently used by industry.

**Fault injection:** Faults possible to simulate have been investigated and identified, and there are no known conflicts with other work packages. As a first step, packet probe blocks have been introduced into the VEP-EX simulation model. Such probes allow dropping or corrupting packets on links between two given DNPs in a custom fashion.

### 3.3.6. AED integration with DNP driver software

During 2011, RWTH and INFN ported DNP device drivers previously written by INFN to the AED. First, the basic DNP RDMA API was ported, which enabled AED software to conduct DNP PUT/GET/SEND operations. After the low-level RDMA API was put into place, the next-higher level Presto communication API has also been ported to the AED, which now enables AED software to conduct synchronous blocking send and receive operations using the DNP 3D torus network. Communication partners are addressed by their respective communication rank in the Presto API.

Further on, being able to run DNP driver code natively on the host using the AED simplifies debugging of such driver code. This is especially true in the event of many cores communicating with each other simultaneously.

AED test applications are currently used as software to further develop the fault-aware DNP model.

### 3.3.7. Hardware-Dependent Software (HdS)

We can split the TIMA contribution and work in 2011 in two main points: port of DNA-OS on top of the target processors and platforms, and specification of research development to reach the overall objective of EURETILE.

Port of DNA-OS: We have ported DNA-OS on top of two new processors: iRiSC (RISC) and x86 (CISC). The port on top of the iRiSC has been started in the last quarter of 2011, after getting development tools from Synopsys for the iRiSC processor, including compiler, linker, simulator, As no architecture has already been provided during 2011, only the kernel of DNA-OS has been ported. C library and peripherals (drivers) are still an on-going work and remain to be done in 2012.

For x86, the kernel has been ported on the x86 architecture with standard peripherals like timers. Standard library is also de facto available after re-writing specific functions (printf,) to fit with peripherals. Moreover, we manage to provide a booting version on USB stick or on x86 simulator (QEMU based) with a simple application using semaphores, multi-threading, timers

Regarding TIMA main contribution on EURETILE, providing task migration on top of NUMA (and non-SMP) architectures, we are looking at virtualization to provide multi-application support as well as other features previously planned for this project. We are following two different tracks: improving DNA-OS to support virtualization, or using an existing virtualization solution (OKL4 provided by NICTA-lab, Sydney, Australia) and adapting it to our requirements. Unfortunately, the OKL4 version we found freely available in 2011 has been an old version that needs an old software tool chain version and it runs only on ARM. As a conclusion, we are heading for a modification of DNA-OS, to provide virtualization support, as well as support for the DAL API described in the deliverable D_WP1_R1. The integration of DAL API for task migration support was still under development at the end of 2011.

## 3.4. Work Performed and Achieved Results in 2012 (Third Year)

### 3.4.1. System-Level Programming Framework (DAL)

During 2012, in the frame of WP3 led by ETHZ, the distributed application layer (DAL) has been finalized and its key concepts published. DAL is a scenario-based design flow that supports design, optimization, and simultaneous execution of multiple applications targeting heterogeneous many-core systems. In particular, the deliverable document produced by WP3 provides detailed insights in the high-level specification model for dynamic systems targeting many-core systems. Afterwards, a multi-objective mapping optimization algorithm is proposed for DAL. As the optimization problems seen in dynamic systems are relatively complex, the proposed technique uses problem decomposition to provide scalability. As many-core systems are prone to high chip temperatures, providing guarantees on maximum temperature is as important as functional correctness and timeliness when designing many-core systems. Thus, two thermal analysis methods for many-core systems are proposed in the WP3 2012 deliverable document. The first method estimates the temperature based on a set of application-specific calibration runs and associated temperature measurements using available built-in sensors. The second method is a formal worst-case real-time analysis method to provide safe bounds on the execution time and the maximum chip temperature. Finally, a hierarchical control mechanism is proposed and it is shown how the hierarchical control mechanism can be used to implement an automatic fault recovery mechanism.

All the discussed concepts have been implemented in a prototype of the distributed application layer. In order to model the high-level specification, DALipse, a software development environment for DAL, has been developed in the frame of WP3. As a result of the 2012 efforts, DALipse is available as a plug-in for Eclipse. In addition, SADEXPO, a state- and architecture-based decomposition framework for mapping optimization is integrated in DALipse. In order to execute applications specified using the DAL paradigm on a distributed Linux cluster, a run-time environment has been implemented using the message-passing interface (MPI) as a communication layer. Finally, a software tool-chain has been developed that uses as input the high-level specification of a DAL benchmark and automatically generates code for the selected target architecture, i.e., a distributed Linux cluster.

### 3.4.2. Hardware-Dependent Software (HdS)

The HdS activity during 2012 could be split in 3 different parts: basic architecture features (for x86-based QUONG architecture and iRiSC-based VEP architecture), the DAL2binary tool chain generation code, and the state of the art of task migration for embedded systems.

We can summarize in few words the main differences between the 2 architectures:
- x86-based QUONG platform: This architecture is based on Intel multi-core x86 architecture. The DNP is accessible through the PCIe interface, as well as peripherals. This supposes the support of PCIe for our own operating system DNA-OS.
- iRiSC based VEP architecture: This architecture is based on the iRiSC processor, with one processor per tile. Each tile includes the DNP, which allows connecting several tiles. There is no PCIe anymore, so the driver for the DNP is directly called by DNA-OS.

For the x86-based QUONG platform, we have done several developments in 2012.
- DNA-OS is now able to boot on multi-core Intel x86 processor.

- The PCIe support is now available in that context.
- We have developed a command interface (as an added DNA-OS service) to facilitate debugging or in-line command in a terminal. This includes a keyboard management.
- We have implemented of the DAL API: start, stop, pause, resume allowing to start, stop, pause and resume task on one processor.
- The support of Ethernet connection has been validated for x86 architecture in DNA-OS. It includes the lightweight IP (open source TCP/IP stack for embedded).
- Interrupt management support by DNA-OS for x86 standard architecture.
- 1 other specific development for the demos: graphical windows on bare metal x86 hardware architecture used by DNA-OS for frame buffer management of video screen.

For the iRiSC-based VEP architecture, we have done one important development in 2012 as well.
- DNA has been ported on iRiSC architecture and validated.
- The DNP driver has been developed and validated on the iRiSC-based architecture. It is worth mentioning that this driver has been first validated on our own ARM-based simulation platform that includes DNP for multi-tile simulation environment.

During 2012, the first version of the DAL2binary tool chain has been provided. This tool chain takes as input the DAL model, and depending on the target, it generates C code that should be compiled with DNA-OS and requested services to get the final binary code for each processor.

Finally, the state of the art regarding the task migration for embedded system has been done as well to fit with the requirements of the EURETILE project depending on the target architecture.

### 3.4.3. Scalable Embedded Computing Simulation Environment (VEP)

During 2012, research activities in simulation and debugging technologies were carried out that improved the VEP environment in terms of efficiency, scalability and usability for the project. The following concrete activities were performed in the simulation ecosystem:

**VEP-EX:** On one hand, the VEP was extended to facilitate the validation of EURETILE's main brain-inspired concepts. The simulator has been provisioned with a service network that completes the necessary artefacts to develop fault monitoring and awareness features across different EURETILE layers (i.e. DNP and HdS). On the other hand, new features to improve the VEP's usability during software development tasks were added. User-configurable loggers and a visualization interface have been developed in order to provide better tracing, analysis and visualization of the software under execution and its output. Finally, some internal components of the VEP-EX were optimized in order to achieve higher simulation speed. Included optimizations help to avoid overhead of unnecessary events inside the SystemC kernel and data transactions in inter-component communication routines. With these new optimizations alone the simulator executes ca. 2.2X faster than the previous version.

**Fault Injection:** A fault injection framework has been created and coupled to the VEP-EX. The framework allows users to inject faults to the system with specific temporal and spatial characteristics, while avoiding the manipulation of complex simulator source code. By using a formalized XML specification of faults, it is possible to force the manifestation of the same faults repeatedly across different simulation executions. This guarantees deterministic faulty scenarios and facilitates the development of fault aware components in different EURETILE layers.

**Parallel Simulation:** The parSC kernel, which was conceived in SHAPES and matured during the previous years of EURETILE, has been used together with the VEP-EX in order to accelerate its execution speed. The use of parSC in EURETILE posed new challenges to guarantee the simulator determinism and its compatibility with legacy sequential models. New technologies were developed (i) to facilitate the integration of legacy models in a practical way and (ii) analyse nondeterministic traits in SystemC code that might lead to functional anomalies (i.e., the SCandal tool). These new technologies were presented in FDL'12 and HLDVT'12, and a joint paper by RWTH, INFN and Synopsys was accepted for publication at VIPES'13. After guaranteeing the simulator determinism, the VEP-EX with parSC was found to achieve a speed-up of ca. 7.77X over the previous version (including the other speed optimizations added to the VEP).

**Debugging and Profiling:** The event-based multi-tile debugger architecture developed in previous years was improved to attain higher retargetability and scalability. Event monitoring in the framework was re-designed as a new component-based system, which facilitates the definition of custom monitors for different OSs and target processors. A well-defined event-based intermediate representation (EIR) was also developed that facilitates the abstraction of complex low-level details of target hardware and OSs into events at a level of abstraction useful for systematic debug. To attain higher scalability, the debugger's structure was organized into a tree-like network of event monitors that allows distributing and parallelizing the debugger itself when dealing with massively parallel systems. Some of the components of the debugger infrastructure have been ported to target the VEP-EX processing components. The new multi-core debugger architecture was presented in S4D'12.
Additionally, a new scriptable, interactive command line interface (CLI) was added to the VEP-EX to allow defining custom automatic debug and profiling tasks. The CLI allows extending basic debug and inspection APIs of the VEP through the TCL programming language.

### 3.4.4. HW support to Fault Awareness

Fault-tolerance issues has been addressed in EURETILE since 2011 with the LO|FA|MO HW/SW approach, whose principles presented in deliverable D6.1 (to be published) have been applied in the DNP SystemC model and lately demonstrated over the VEP simulator at the 2012 review meeting (activity reported by D5.2).
The "fault awareness" activities in 2012 focused on the implementation of LO|FA|MO mechanism for the EURETILE HPC experimental platform, and resulted in a consolidated specification of the LO|FA|MO approach, the integration of the VHDL code of DNP Fault Manager in APEnet+ hardware, the introduction of a controller of the diagnostic messages flow over the 3D Torus Network (the Link Fault Manager, LiFaMa), and the first implementation of the Host Fault Manager as a Linux daemon. In the current implementation the Nios II micro-controller performs the reading process of the FPGA sensors over the Avalon Bus and we foresee to implement a technique to transfer the information to the DNP status registers. The new functional blocks were also verified through an extensive test session.

### 3.4.5. DNP and APENet+ Innovations

During 2012, the integration of the complete DNP IP in APEnet+ hardware system allowed real tests of APEnet+ architecture in the final environment.

First, INFN released a library of test programs based on synthetic and real application kernels and performed extensive test sessions to measure bandwidth and latency in different conditions.

Benchmark results were used to optimize specific hardware blocks impacting on system performances. In particular we executed some re-design activities on 3D links sensible obtaining channel bandwidth enhancement and a more robust transfer protocol.

Further activities were also devoted to optimization of the APEnet+ host interface where the PCIe protocol shows high latency that has to be controlled and minimized: the Host Interface TX block has been completely re-designed and the software interface accelerated moving some critical tasks to FPGA custom hardware blocks. In cooperation with TARGET, preliminary evaluation of effects of introduction on RX RDMA path of a TLB (Translation Look aside Buffer) for virtual to physical address translation has been performed.

The P2P GPU interface was improved and heavily tested achieving good results in terms of bandwidth enhancement and small packets latency reduction.

In order to explore future systems development activities in 2012 we executed the porting of our DNP IP design on state-of-the-art FPGA devices that, built using a 28nm silicon process, show impressive amount of internal user-available resources and huge number of high speed transceivers. The preliminary evaluation and synthesis activities showed very encouraging results: the DNP logic occupation is limited to 15% of a medium size devices (allowing architecture improvement through introduction of new computing/specialized hardware) while its I/O throughput can double thanks to the use of the 28nm devices improved transceivers.

Software activities focused on system software bug fixing, performance optimization and development of new modules supporting hardware innovations.

Highlights here are the release of a VEP interface for Presto programming environment and the development of APEnet+ API OpenMPI layer to support DAL on APEnet+ platform. Furthermore, in collaboration with UJF, a DNA-OS prototype driver specific to the APEnet+ hardware has been analysed and a simple program produced by the DNA-OS tool chain, running on x86 and setting selected APEnet+ board registers through the PCIe was coded.

Several exploitation activities are also in place from 2012. In particular we demonstrated the advantages of DNP/APEnet+ architecture introduction in high-level trigger systems of the current and future High Energy Physics colliders. In this framework the APEnet+ features, one for all the GPU Direct P2P mechanism assuring low access latency to numerical accelerators, and its implementation FPGA-based allowing IP re-configurability and specialization, can reduce data transport overheads and can increase the time budget for trigger computing.

### 3.4.6. ASIP support to RDMA

EURETILE's many-core platforms critically depend on efficient packet-based communication between tiles in the 3D network. Efficiency refers to both high bandwidth and low latency. Our observation for the HPC platform is that the current DNP implementation, which includes firmware running on a NIOS soft-core from Altera, offers high-bandwidth communication, but with a too high latency. To reduce the latency, a significant acceleration of networking functions (e.g. buffer search, virtual-to-physical address translation) is needed.

During 2012, TARGET in cooperation with INFN started the development of a software-programmable DNP architecture using ASIP technology. These activities are part of WP6. A first version of a DNP ASIP has been modelled and optimised, using the IP Designer tool-suite. The basic architecture is a 32-bit microprocessor. Adding dedicated hardware and instructions to accelerate the buffer-search functionality for RDMA tasks has further optimised the architecture. As a result, the cycle count of typical buffer-search firmware programs has been reduced by more than 85% compared to the existing implementation on NIOS. Currently the clock frequency of the ASIP is still lower than that of NIOS. Nonetheless, buffer search already executes significantly faster on the ASIP compared to NIOS, and moreover further cycle-time improvements are within reach.

### 3.4.7. ASIP Design Tool Improvement.

EURETILE's approach towards the design of ASIPs is to use a retargetable tool-suite based on a processor description language. The tool-suite supports architectural exploration, the generation of a software development kit, and the generation of an RTL hardware implementation of the ASIP. TARGET's IP Designer tool-suite is used for this purpose.

During 2012, TARGET developed new extensions and optimisations for the retargetable tool-suite, with specific relevance to the other Euretile activities. These tool development activities are part of WP6. In particular, we focussed on the following topics:

- We continued the work on the development of advanced capabilities for modelling and synthesis of I/O interfaces of ASIPs. While the basic concepts behind this new approach were already reported last year, this year we have delivered a first implementation of the new concepts, and tested it by modelling and synthesising first example I/O interfaces.

- The tools have been extended with capabilities to model multi-threaded architectures. This includes the generation of hardware support for storing the context of a thread upon a context switch, as well as C programming aspects.

Finally we developed and implemented additional capabilities to provide feedback to users of the retargetable tool-suite during the ASIP architectural optimisation process. Specifically, the tools can now analyse and report about connectivity issues within an ASIP architecture.

### 3.4.8. HW Experimental Platform Prototype (QUonG)

During 2012, INFN completed the procurement and assembly phase of a 16 nodes QUonG system, a 16K core, 32 TFlops peak single precision computing platform with a 4x2x2 topology. All required servers and GPU accelerators were procured from H1-2012 while a batch of 15 APEnet+ boards were delivered in September 2012 and put in

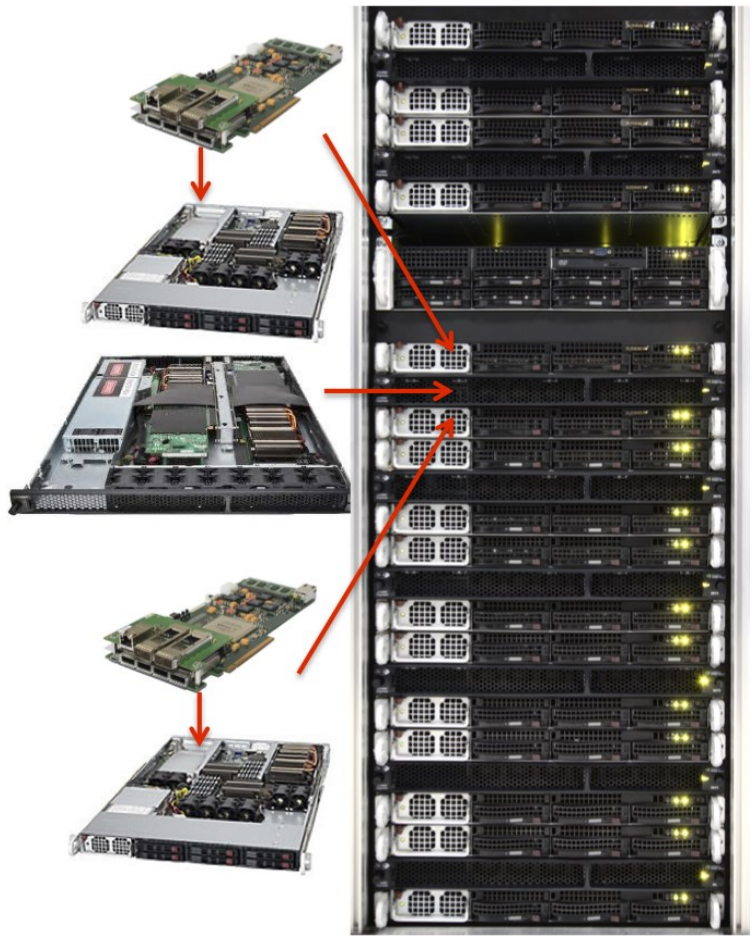

**Figure 3-7. 16 nodes QUonG prototype in 2012**

place in H2-2012.

Furthermore, a 12 APEnet+ boards addictive production was launched in H2-2012 and the boards will be delivered in February 2013.

In parallel a reduced QUonG platform prototype (8 nodes, 2x2x2 topology) was heavily tested to validate systems hardware and software and to provide a prototype system to develop applications.

In addition to the EURETILE scientific applications and benchmarks, the QUonG prototype (with and without APEnet+ network) was validated through the successful execution of several grand challenges application kernels among them the Heisenberg Spin Glass code (HSG), the Graph500 benchmark, Laser-Plasma Interaction simulation and preliminary HEP trigger related algorithms.

### 3.4.9. DPSNN-STDP: Distributed Polychronous and Plastic Spiking Neural Net

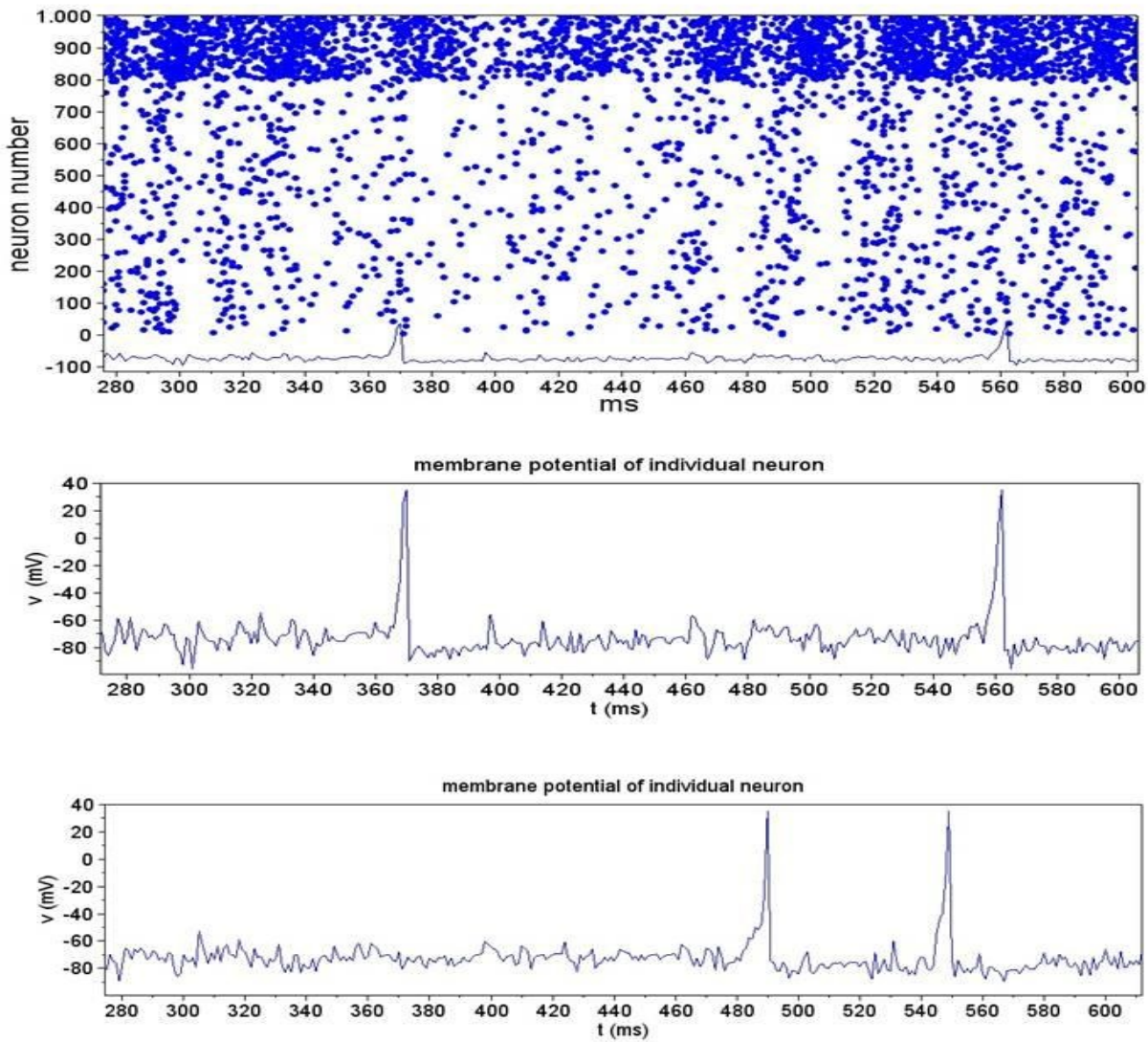

**Figure 3-8. Collective spiking rastergram and individual neuron activity produced by DPSNN-STDP.**

During 2012, we designed and implemented the C++ code of a complete prototype of the DPSNN-STDP simulator (Distributed Polychronous Spiking Neural Net with synaptic Spiking Time Dependent Plasticity).

Then we run the full simulation cycle (initialization and dynamic phases of a network including 10^5 synapses) in the C++ plus MPI environment on a QUONG prototype, using an industry standard interconnection system (InfiniBand switched fabric), reproducing an identical behavior with different number of processes.

We ported and run the initialization phase of the DPSNN-STDP code under the DAL plus C++ environment, and executed it on a prototype environment where two QUONG nodes were interconnected by two APENet+ card. We used the same C++ classes as building blocks to construct the DAL and MPI network of processes. This will permit, during 2013, to start a comparison activity between the APENet+ interconnect system and the InfiniBand system, and among the performances offered by the DNA-OS, MPI and Presto message passing software layers, when applied to the DPSNN-STDP benchmark. Then we will identify the middleware and hardware bottlenecks, and derive architectural improvements for future generations of interconnect systems.

## 3.5. Plans for 2013 - 2014 activities

The prototypes of the Experimental hardware platform (QUonG), of the simulator (VEP), and of the integrated software tool-chain based on the DAL programming environment and on the EURETILE HdS should be delivered before the end of 2013. By the end of the 2013, each work-package will have demonstrated working prototypes about fault management, description of application dynamism and support of scalability. During 2013, the consortium will elaborate a more complete picture about those subjects and, taking as example the fault management area, reduce the distance between what each partner is doing to detect and manage faults and what is actually done at integrated level. By the end of 2013, we plan to have completed the porting of the full DPSNN-STDP cortical simulator on the EURETILE platform and started a campaign of measurements to identify the most promising areas of architectural improvements. A possible extension of the project end until September 2014 is under evaluation. At project level, we foresee a few main topics that should be covered during 2014:

- **Qualitative assessments and quantitative measurements.** The availability of the integrated EURETILE system would enable a more extensive campaign of qualitative evaluations and quantitative measurements of our integrated software and hardware platforms and development environment. This way, the consortium could elaborate about the main EURETILE merits and further developments that we perceive necessary on the bases of an adequate temporal period dedicated to the evaluation of the platform developed.
- **Better integration of management's methods of fault, application dynamism and scalability.** During 2014, the existing demonstrators would be improved toward greater system integration. We will also develop benchmarks dedicated to show application dynamism and use the benchmarks to highlight the fault-tolerant behaviours.
- **Brain-Inspired enhancements.** During 2014, we will start experiments about hardware and software improvements derived from the experience gained thanks to the run of the DPSNN-STDP benchmark.

## 3.6. **The SHAPES Project Background (2006-2009)**

As the EURETILE project is an evolution of the SHAPES FP7 project, we insert here a summary of the features of that project that has been relevant for the EURETILE project.

### 3.6.1. **2006 SHAPES Project Abstract**

There is no processing power ceiling for low consumption, low cost, dense Numerical Embedded Scalable Systems dedicated to future human-centric applications, which will manage multi-channel audio, video and multi-sensorial input/outputs.

Nanoscale systems on chip will integrate billion-gate designs. The challenge is to find a scalable HW/SW design style for future CMOS technologies.

The main HW problem is wiring, which threatens Moore's law. Tiled architectures suggest a possible HW path:

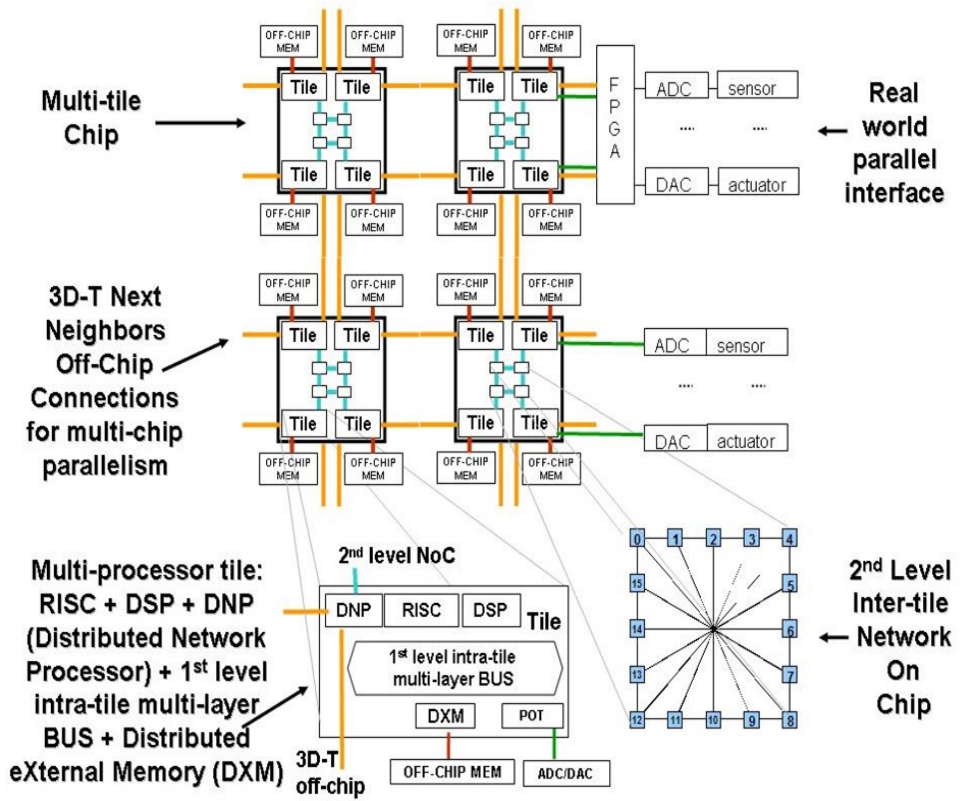

"small" processing tiles connected by "short wires". A second HW problem is the management of the design complexity of billion gate designs. A tiled design style extensively reuses processing tiles, each tile composed of stable Intellectual Properties requiring only a few million gates: a manageable complexity. The SW challenge is to provide a simple and efficient programming environment for a (massive) tiled parallel architecture. The proposed approach must be experimented through applications.

**Figure 3-9. SHAPES Architecture template**

### 3.6.2. **Investigation of the tiled HW paradigm**

SHAPES investigated a ground-breaking HW/SW architecture paradigm. A typical heterogeneous SHAPES tile is composed of a VLIW floating-point DSP (Digital Signal Processor), a RISC controller, a DNP (Distributed Network Processor), distributed on chip memory, a POT (a set of Peripherals On Tile) plus an interface for DXM (Distributed External Memory). Each Tile includes a few million gates, for optimal balance among parallelism, local memory, and IP reuse on future technologies.

The SHAPES routing fabric connects on-chip and off-chip tiles, weaving a distributed packet switching network. 3D next-neighbours engineering methodologies have been studied for off-chip networking and maximum system density, see the Figure 3 on previous page. SHAPES opened the path towards new density records, with multi-Teraflops single-board Tiled computers and Petaflops systems.

### 3.6.3. Experimentation of real-time, communication aware system SW

SHAPES adopted layered system software, which does not destroy the information about algorithmic parallelism, data and workload distribution and real-time requirements provided by the programmer.

The system software should be fully aware of the tiled hardware paradigm. For efficiency and predictability, the system software manages intra-tile and inter-tile latencies, bandwidths, computing resources, using application- and architecture-level profiling.

(a) Application description using the Process Network formalism and a simplified representation of the SHAPES tile. (b) Mapping of the application process network on the SHAPES tile

The application is described, using a model based approach, in terms of a network of actors, with explicit real-time constraints. Figure 4a shows an illustration of the application description). The application is mapped to the SHAPES architecture by an iterative automated (or semi-automated) multi-objective optimization. The mapping procedure uses performance estimations obtained at different levels of abstraction, i.e. system-level analytic predictions based on results of the low-level simulation environment. Figure 4b shows a mapping representation.

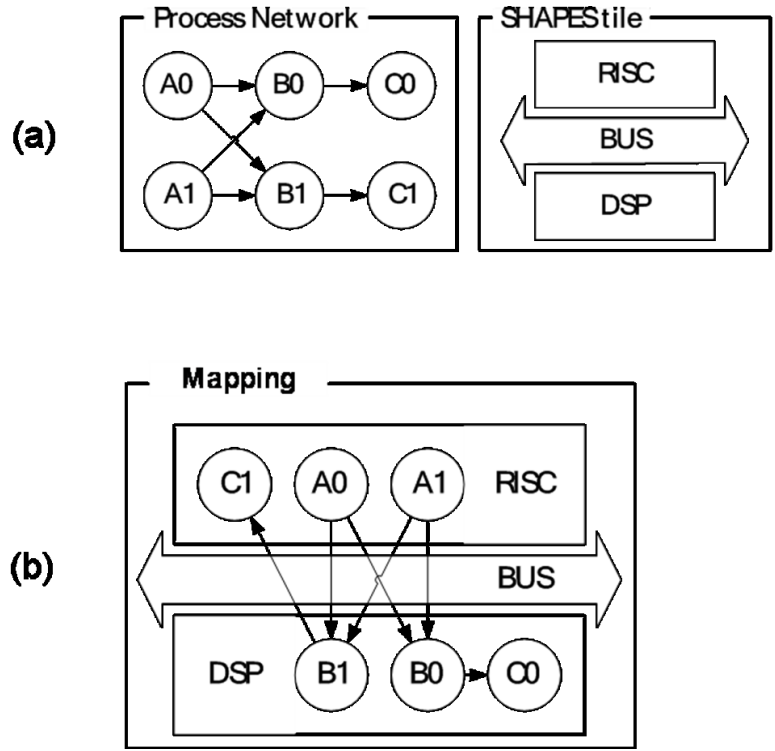

**Figure 3-10. (a) Application description using the Process Network formalism and a sumplified representation of the SHAPES tile. (b) Mapping og the application process network onthe SHAPES tile.**

The layered structure of the software separates the application code from the Hardware-dependent Software (HdS). Therefore it is possible to debug the application and system software layers independently and obtain higher simulation speeds. The application can be debugged using a virtual architecture, which is agnostic of the HdS details and the real hardware properties. The HdS generator refines the system software, first through a transaction-accurate step, and then down to the virtual prototype level (i.e. using the virtual Shapes platform – VSP), where the generated system software contains all the details needed in order to be executed on the actual hardware platform.Figure 5 shows the different components of the SHAPES SW environment as well as their interactions. In EURETILE, this concept of the SW environment, the interfaces between the various tools and the software design flow will be adapted towards the new challenges imposed by the three-layer hierarchy and the dynamic adaptation capabilities.

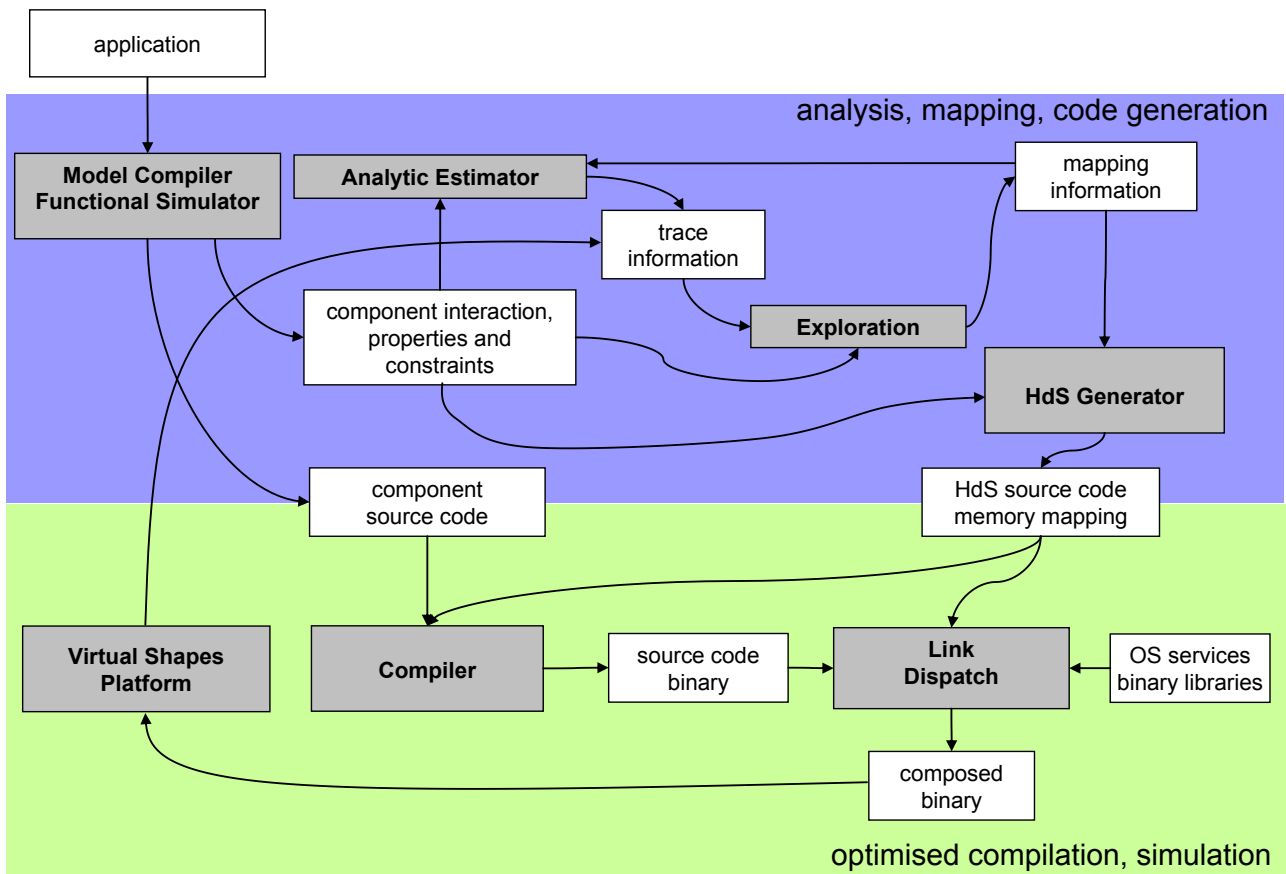

**Figure 3-11. SHAPES software environment: overview.**

The software accesses on-chip and off-chip networks through a homogeneous interface. The same hardware and software interface can be adopted for integration with signal acquisition and reconfigurable logic tiles. Generation after generation, the number of tiles on a single-chip will grow, but the application will be portable.

The SHAPES HW and SW platform has been benchmarked through a set of applications characterized by a large inherent parallelism and, with one exception, by real-time constraints: wave field synthesis for array of sound sources reproduced by large arrays of loud-speakers, treatment of audio signals acquired by arrays of microphones, Ultrasound Scanners, and simulation of Theoretical Physics (Lattice Quantum-Chromo Dynamics).

SHAPES proposed a programmable, high performance, low power, dense system solution designed for interfacing with reconfigurable logic and signal acquisition and generation systems. The architecture is designed to scale from:

- low end single module hosting 1-8 tiles for mass market applications
- classic digital signal processing systems like radar and medical equipment (2 K tiles)
- high-end systems requiring massive numerical computation (32 K tiles)

### 3.6.4. DIOPSIS940HF MPSOC and SHAPES RISC + mAgic VLIW DSP + DNP Tile.

ATMEL ROMA, partner of the SHAPES project, delivered (2008) the DIOPSIS940HF 130 nm CMOS silicon (RISC + mAgicV VLIW DSP MPSoC) and development board, which constituted the HW reference model for the SW simulator of the computational tile and led place & route trials on advanced technology (65 and/or 45 nm) to assess the scalability toward HW multi-tiling. During 2009, the consortium finalized the tape-out trial of an eight tiles chip using a 45 nm CMOS target technology, this way getting final (for inter-tile communications), a floating-point VLIW processor (for numerical intensive computations) and a RISC processor (for control, user interface and sequential computations).

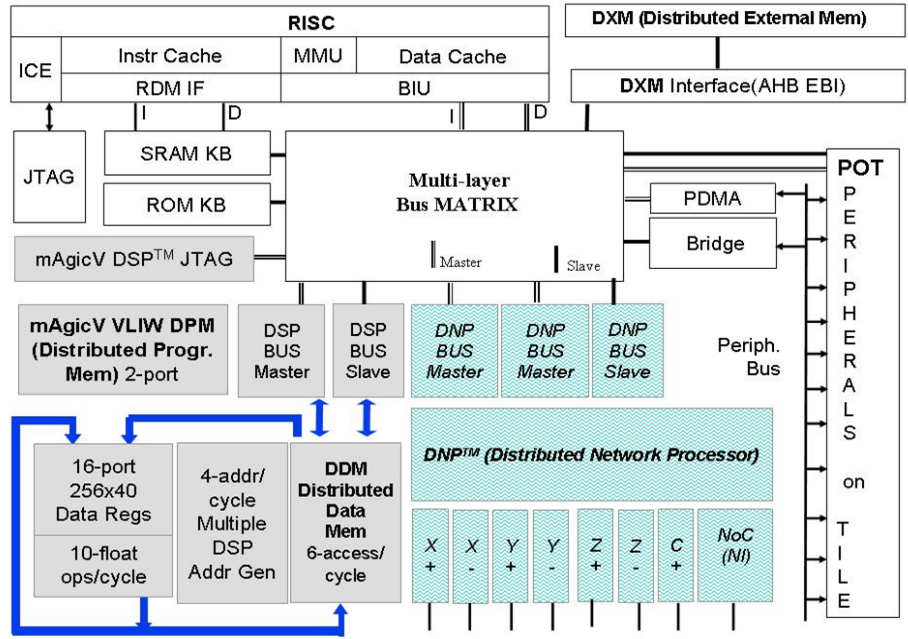

**Figure 3-14. The SHAPES elementary tile: RISC + VLIW DSP + Distributed Network Processor.**

figures for Area (56 mm$^2$), Computational Power (25 GigaFlops), and Power Consumption (2.9 W). Each elementary hardware tile is a multi-processor, which includes a distributed network processor

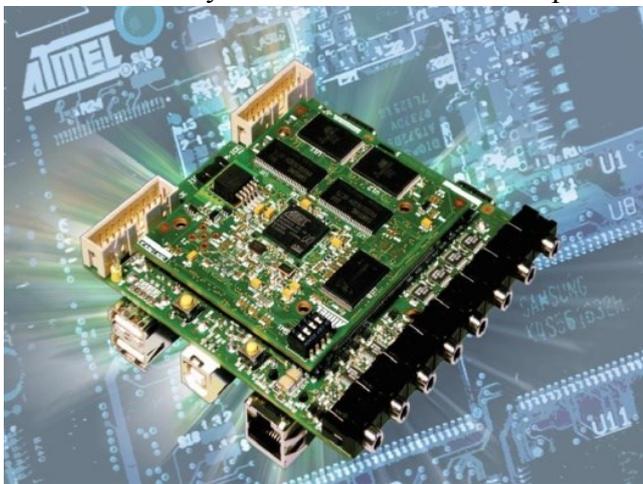

**Figure 3-13. Board mounting the DIOPSIS 940 HF RISC + mAgic VLIW DSP multi-processor.**

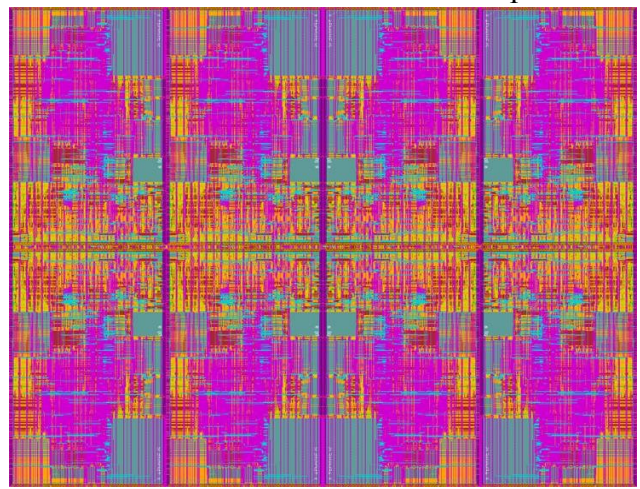

Figure 3-12. Eight SHAPES tiles placed&routed in 45 nm.

The SHAPES Hardware tile was a joint design effort of the APE team of INFN (the Italian Istituto Nazionale di Fisica Nucleare) and ATMEL Roma (here, we take to opportunity to acknowledge the work of Federico Aglietti, Antonio Cerruto, Antonio Costa, Maurizio Cosimi, Andrea Michelotti,

Elena Pastorelli, Andrea Ricciardi and of all the designers of the DIOPSIS design center). University of Roma (Prof. Alessandro Trifiletti and Giuseppe Scotti, among the others) played a key role in the place & route on the target 45 nm CMOS technology of the 8 tiles SHAPES chip shown in the picture. ST Microelectronics (Marcello Coppola and his team), University of Cagliari (prof. Luigi Raffo and his team) and Pisa (Prof. Luca Fanucci and his team), evolved the Spidergon Network-on-Chip to be used for inter-tile communication, attached to the Distributed Network Processor designed by INFN.

# 4. Dissemination

During this period, the EURETILE project produced:
- A set of publications and presentations describing the EURETILE results, listed by the first sections of this chapter;
- The CASTNESS'11 workshop (Roma, 17-18 January 2011). Twenty speakers, representing the four Teradevice Computing projects (EURETILE, TERAFLUX, TRAMS and SooS) presented their activities to about 60 researchers. The last section of this document details the lists of presentations and participants.

## 4.1. **Papers - Posters - Technical Press – Newsletters**

**ETHZ**

- P. Kumar, and L. Thiele. Behavioural Composition: Constructively Built Server Algorithms. Proc. 5th Workshop on Compositional Theory and Technology for Real-Time Embedded Systems, San Juan, Puerto Rico, p. 9-12, Dec. 2012.
- P. Kumar and L. Thiele. Quantifying the Effect of Rare Timing Events with Settling-Time and Overshoot. Proc. IEEE Real-Time Systems Symposium (RTSS), San Juan, Puerto Rico, p. 149-160, Dec. 2012.
- S.-H. Kang, H. Yang, L. Schor, I. Bacivarov, S. Ha, and L. Thiele. Multi-Objective Mapping Optimization via Problem Decomposition for Many-Core Systems. Proc. IEEE Symposium on Embedded Systems for Real-Time Multimedia (ESTIMedia), Tampere, Finland, p. 28-37, Oct. 2012.
- L. Schor, I. Bacivarov, D. Rai, H. Yang, S.-H. Kang, and L. Thiele. Scenario-Based Design Flow for Mapping Streaming Applications onto On-Chip Many-Core Systems. Proc. Int'l Conf. on Compilers Architecture and Synthesis for Embedded Systems (CASES), Tampere, Finland, p. 71-80, Oct. 2012.
- D. Rai, H. Yang, I. Bacivarov, and L. Thiele. Power Agnostic Technique for Efficient Temperature Estimation of Multicore Embedded Systems. Proc. Int'l Conf. on Compilers Architecture and Synthesis for Embedded Systems (CASES), Tampere, Finland, p. 61-70, Oct. 2012.
- L. Schor, H. Yang, I. Bacivarov, and L. Thiele. Thermal-Aware Task Assignment for Real-Time Applications on Multi-Core Systems. Proc. Int'l Symposium on Formal Methods for Components and Objects (FMCO) 2011, Turin, Italy, Volume 7542 of LNCS, p. 294-313, Oct. 2012.
- L. Schor, H. Yang, I. Bacivarov, and L. Thiele. Worst-Case Temperature Analysis for Different Resource Models. IET Circuits, Devices & Systems, Volume 6, Issue 5, p. 297-307, Sep. 2012.
- K. Huang, W. Haid, I. Bacivarov, M. Keller, L. Thiele. Embedding Formal Performance Analysis into the Design Cycle of MPSoCs for Real-time Streaming Applications. ACM Transactions in Embedded Computing Systems (TECS Journal), ACM, Volume 11, Issue 1, p. 8:1-8:23, 2012.
- L. Schor, I. Bacivarov, H. Yang, and L. Thiele. Fast Worst-Case Peak Temperature Evaluation for Real-Time Applications on Multi-Core Systems. Proc. IEEE Latin American Test Workshop (LATW), Quito, Ecuador, p. 1-6, Apr. 2012.
- L. Schor, I. Bacivarov, H. Yang, and L. Thiele. Worst-Case Temperature Guarantees for Real-Time Applications on Multi-Core Systems. Proc. IEEE Real-Time and Embedded Technology and Applications Symposium (RTAS), Proc. IEEE Real-Time and Embedded Technology and Applications Symposium (RTAS), Beijing, China, p. 87-96, Apr. 2012.
- P. Kumar and L. Thiele. Timing Analysis on a Processor with Temperature-Controlled Speed Scaling. Proc. IEEE Real-Time and Embedded Technology and Applications Symposium (RTAS), Beijing, China, p. 77-86, Apr. 2012.
- I. Bacivarov, I. Belaid, A. Biagioni, A. El Antably, N. Fournel, O. Frezza, J. Jovic, R. Leupers, F. Lo Cicero, A. Lonardo, L. Murillo, P.S. Paolucci, D. Rai, D. Rossetti, F. Rousseau, L. Schor, C. Schumacher, F. Simula, L. Thiele, L. Tosoratto, P. Vicini, H. Yang – "DAL: Programming Efficient and Fault-Tolerant Applications for Many-Core Systems" - Poster at HIPEAC12 - Jan 23-25, 2012 Paris, France

- P. Kumar, J.-J. Chen, and L. Thiele. Demand Bound Server: Generalized Resource Reservation for Hard Real-Time Systems. Proc. Int'l Conference on Embedded Software (EMSOFT), pages 233-242, Oct. 2011.
- L. Schor, H. Yang, I. Bacivarov, and L. Thiele. Worst-Case Temperature Analysis for Different Resource Availabilities: A Case Study. Proc. Workshop on Power and Timing Modeling, Optimization and Simulation (PATMOS), Lecture Notes on Computer Science (LNCS), Springer, Vol. 6951, pages 288-297, Sep. 2011.
- P. Kumar, J.-J. Chen, L. Thiele, A. Schranzhofer, and G. C. Buttazzo. Real-Time Analysis of Servers for General Job Arrivals. Proc. Intl. Conf. on Embedded and Real-Time Computing Systems and Applications (RTCSA), pages 251-258, Aug. 2011.
- L. Thiele, L. Schor, H. Yang, and I. Bacivarov. Thermal-Aware System Analysis and Software Synthesis for Embedded Multi-Processors. Proc. Design Automation Conference (DAC), pages 268-273, Jun. 2011.
- D. Rai, H. Yang, I. Bacivarov, JJ. Chen, L. Thiele. Worst-Case Temperature Analysis for Real-Time Systems. In Proceedings of Design, Automation and Test in Europe (DATE), Grenoble, France, March 2011.
- S. Perathoner, K. Lampka, L. Thiele. Composing Heterogeneous Components for System-wide Performance Analysis. In Proceedings of Design, Automation and Test in Europe (DATE), Grenoble, France, March 2011 (invited paper).
- I. Bacivarov, H. Yang, L. Schor, D. Rai, S. Jha, L. Thiele, Poster: Distributed Application Layer - Towards Efficient and Reliable Programming of Many-Tile Architectures. Design, Automation and Test in Europe (DATE) Friday Workshop, Grenoble, France, March 2011.
- K. Huang, L. Santinelli, JJ. Chen, L. Thiele, and G. C. Buttazzo. Applying Real-Time Interface and Calculus for Dynamic Power Management in Hard Real-Time Systems. Real-Time Systems Journal, Springer Netherlands, Vol. 47, No. 2, pages 163-193, Mar. 2011.
- A. Schranzhofer, JJ. Chen, L. Thiele. Dynamic Power-Aware Mapping of Applications onto Heterogeneous MPSoC Platforms. IEEE Transactions on Industrial Informatics, IEEE, Vol. 6, No. 4, pages 692 -707, November, 2011.

**RWTH**

- C. Schumacher, J. H. Weinstock, R. Leupers, G. Ascheid, L. Tossorato, A. Lonardo, D. Petras, T. Groetker. "legaSCi: Legacy SystemC Model Integration into Parallel SystemC Simulators". 1st Workshop on Virtual Prototyping of Parallel and Embedded Systems (ViPES), 2013, Boston, USA (accepted for publication).
- C. Schumacher, J. H. Weinstock, R. Leupers and G. Ascheid: Cause and effect of nondeterministic behavior in sequential and parallel SystemC simulators. IEEE International High Level Design Validation and Test Workshop (HLDVT'12). Nov 2012, Huntington Beach (California-USA).
- C. Schumacher, J. H. Weinstock, R. Leupers and G. Ascheid: Scandal: SystemC Analysis for NonDeterminism AnomaLies. Forum on Specification and Design Languages (FDL '12), Sep 2012, Vienna (Austria)
- L. G. Murillo, J. Harnath, R. Leupers and G. Ascheid: Scalable and Retargetable Debugger Architecture for Heterogeneous MPSoCs. System, Software, SoC and Silicon Debug Conference (S4D '12), Sep 2012, Vienna (Austria)
- L. G. Murillo, J. Eusse, J. Jovic, S. Yakoushkin, R. Leupers and G. Ascheid: Synchronization for Hybrid MPSoC Full-System Simulation. Design Automation Conference (DAC '12), Jun 2012, San Francisco (USA)
- R. Leupers: More Real Value for Virtual Platforms. Design, Automation and Test in Europe (DATE '12), Mar 2012, Dresden (Germany)
- J. Jovic, S. Yakoushkin, L. G. Murillo, J. Eusse, R. Leupers and G. Ascheid: Hybrid Simulation for Extensible Processor Cores. Design, Automation and Test in Europe (DATE '12), Mar 2012, Dresden (Germany)
- S. Kraemer, R. Leupers, D. Petras, T. Philipp, A. Hoffmann: Checkpointing SystemC-Based Virtual Platforms. International Journal of Embedded and Real-Time Communication Systems (IJERTCS), vol. 2, no. 4, 2011
- L. G. Murillo, W. Zhou, J. Eusse, R. Leupers, G. Ascheid: Debugging Concurrent MPSoC Software with Bug Pattern Descriptions. System, Software, SoC and Silicon Debug Conference (S4D '11), Oct 2011, Munich (Germany)
- R. Leupers, G. Martin, N. Topham, L. Eeckhout, F. Schirrmeister, X. Chen: Virtual Manycore Platforms: Moving Towards 100+ Processor Cores. Design Automation & Test in Europe (DATE), Mar 2011, Grenoble (France)
- J. Castrillon, A. Shah, L. G. Murillo, R. Leupers, G. Ascheid: Backend for Virtual Platforms with Hardware Scheduler in the MAPS Framework. 2nd IEEE Latin America Symp. on Circuits and Systems, Feb 2011, Bogota (Colombia)
- S. Kraemer, Design and analysis of efficient MPSoC simulation techniques, dissertation, 2011, Aachen, Germany (http://darwin.bth.rwth-aachen.de/opus3/frontdoor.php?source_opus=3769&la=de)

- C. Schumacher, R. Leupers, D. Petras and A. Hoffmann. parSC: Synchronous Parallel SystemC Simulation on Multi-Core Host Architectures. In proceedings of CODES/ISSS '10, October, 2010, Scottsdale, Arizona, USA (http://dx.doi.org/10.1145/1878961.1879005)

**TIMA**
- A. Chagoya-Garzon, F. Rousseau, F. Pétrot, Multi-Device Driver Synthesis Flow for Heterogeneous Hierarchical Systems, Euromicro Conference on Digital System Design, Sept 2012, pp. 389 – 396, Izmir, Turkey.
- Ashraf Elantably, Frédéric Rousseau, Task migration in multi-tiled MPSoC: Challenges, state-of-the-art and preliminary solutions, Journée National du Réseau Doctoral en Microélectronique, Marseille, France, June 2012, Poster and 4 pages paper (in English)
- Chagoya-Garzon, N. Poste, F. Rousseau, Semi-Automation of Configuration Files Generation for Heterogeneous Multi-Tile Systems, Computer Software and Application Conference (COMPSAC 2011), Munich, Germany, 18-21 July 2011.
- H. Chen, G. Godet-Bar, F. Rousseau, F. Petrot, Me3D : A Model-driven Methodology expediting Embedded Device Driver Development, International Symposium on Rapid System Prototyping (IEEE RSP 2011), pp. 171-177, May 2011, Karlsruhe, Germany.

**INFN**
- Roberto Ammendola, Andrea Biagioni, Ottorino Frezza, Francesca Lo Cicero, Alessandro Lonardo, Pier Stanislao Paolucci, Davide Rossetti, Francesco Simula, Laura Tosoratto, Piero Vicini - QUonG: A GPU-based HPC System Dedicated to LQCD Computing - Application Accelerators in High-Performance Computing, Symposium on, pp. 113-122, 2011 Symposium on Application Accelerators in High-Performance Computing, 2011 [[1]]
- Pier Stanislao Paolucci - FP7 EURETILE Project: EUropean REference TILed architecture Experiment - HipeacInfo, Quarterly Newsletter, Number 24, page 11, October 2010 (http://www.Hipeac.net/newsletter) File:PaolucciEuretileHipeacInfo24October2010.pdf
- Roberto Ammendola, Andrea Biagioni, Ottorino Frezza, Francesca Lo Cicero, Alessandro Lonardo, Pier Paolucci, Roberto Petronzio, Davide Rossetti, Andrea Salamon, Gaetano Salina, Francesco Simula, Nazario Tantalo, Laura Tosoratto, Piero Vicini - APEnet+: a 3D toroidal network enabling Petaflops scale Lattice QCD simulations on commodity clusters - High Energy Physics - Lattice 2010 (hep-lat); Distributed, Parallel, and Cluster Computing (cs.DC) arXiv:1012.0253v1 [hep-lat] (proceedings)
- R. Ammendola, A. Biagioni, O. Frezza, F. Lo Cicero, A. Lonardo, P.S. Paolucci, D. Rossetti, A. Salamon, G. Salina, F. Simula, L. Tosoratto, P. Vicini - apeNET+: High Bandwidth 3D Torus Direct Network for PetaFLOPS Scale Commodity Clusters, International Conference on Computing in High Energy and Nuclear Physics (CHEP), October 2010, Taipei, Taiwan - Proceedings on J. Phys.: Conf. Ser. 331 052029 doi:10.1088/1742-6596/331/5/052029
- R. Ammendola, A. Biagioni, O. Frezza, F. Lo Cicero, A. Lonardo, P.S. Paolucci, D. Rossetti, A. Salamon, G. Salina, F. Simula, L. Tosoratto, P. Vicini - Mastering multi-GPU computing on a torus network - GPU Technology Conference 2010 (GTC) - http://www.nvidia.com/content/GTC/posters/2010/I09-Mastering-Multi-GPU-Computing-on-a-Torus-Networki.pdf (poster)
- R. Ammendola, A. Biagioni, G. Chiodi, O. Frezza, A.Lonardo, F. Lo Cicero, R. Lunadei, D. Rossetti, A. Salamon, G. Salina, F. Simula, L. Tosoratto, P. Vicini - High speed data transfer with FPGAs and QSFP+ modules - Topical Workshop on Electronics for Particle Physics, Aachen, Germany / September 20-24, 2010 - Proceedings on JINST 5 C12019 doi:10.1088/1748-0221/5/12/C12019
- R. Ammendola et al. - High speed data transfer with FPGAs and QSFP+ modules - Nuclear Science Symposium Conference Record (NSS/MIC) 2010 IEEE, Publication Year: 2010, Page(s): 1323 1325, November 2010, Knoxville, Tennessee. DOI: 10.1109/NSSMIC.2010.5873983

## 4.2. Books/Book Chapters

**ETHZ**
- Book chapter: I. Bacivarov, W. Haid, K. Huang, L. Thiele. Methods and Tools for Mapping Process Networks onto Multi-Processor Systems-On-Chip. Handbook of Signal Processing Systems, Springer, pages 1007-1040, October, 2010.

**RWTH**
- Book: R. Leupers and O. Temam (Eds.), Processor and System-On-Chip Simulation, Springer, September 2010, ISBN 978-1441961747
- Book: T. Kempf, G. Ascheid, R. Leupers: Multiprocessor Systems on Chip: Design Space Exploration, Springer, Feb 2011, ISBN 978-1441981523

**TIMA**
- Katalin Popovici, Frederic Rousseau, Ahmed A. Jerraya, Marilyn Wolf: Embedded Software Design and Programming of Multiprocessor System-on-Chip, Simulink and SystemC Case Studies, Springer, April 2010, ISBN 978-1-4419-5566-1
- Book chapter: Xavier Guerin, Frederic Petrot, Operating System Support for Applications targeting Heterogeneous Multi-Core System)on-Chip in the book Multi-Core Embedded Systems, CRC Press, Chapter 9, 24 pages, April 2010.

## 4.3. Presentations

**ETHZ**
- Iuliana Bacivarov, Thermal-Aware Design of Real-Time Multi-Core Embedded Systems, Invited talk at Mapping Applications to MPSoCs, Jun. 2011.
- Iuliana Bacivarov, Temperature Predictability in Multi-Core Real-Time Systems, Invited talk at DAC Workshop on Multiprocessor System-on-Chip for Cyber Physical Systems: Programmability, Run-Time Support, and Hardware Platforms for High Performance Embedded Applications, Jun. 2011.
- Iuliana Bacivarov, Distributed Application Layer – Towards Seamless Programming of Many-Tile Architectures, CASTNESS 2011, 17 and 18 January 2011, Rome, Italy,http://euretile.roma1.infn.it/mediawiki/img_auth.php/8/88/EURETILE-2-IulianaBacivarov.pdf.
- Iuliana Bacivarov, Distributed Operation Layer: An Efficient and Predictable KPN-Based Design Flow, invited talk at Workshop on Compiler-Assisted System-On-Chip Assembly 2010, in conjunction with Embedded Systems Week, Scottsdale, AZ, US, October 2010, http://www12.cs.fau.de/ws/casa10.
- Iuliana Bacivarov, Efficient Execution of Kahn Process Networks on CELL BE, invited talk at Summer School on Models for Embedded Signal Processing Systems at Lorentz Center, Leiden, Netherlands, 30 Aug - 3 Sep 2010, http://www.lorentzcenter.nl/lc/web/2010/427/presentations/Iuliana-cell.pdf.
- Iuliana Bacivarov, Distributed Operation Layer: A Practical Perspective, tutorial at Summer School on Models for Embedded Signal Processing Systems at Lorentz Center, Leiden, Netherlands, 30 Aug - 3 Sep 2010, http://www.lorentzcenter.nl/lc/web/2010/427/presentations/Iuliana-demo.pdf.
- Iuliana Bacivarov, Distributed Operation Layer: Efficient Design Space Exploration of Scalable MPSoC, invited talk at Combinatorial Optimization for Embedded System Design workshop 2010 in conjunction with CPAIOR2010, 7th International Conference on Integration of Artificial Intelligence and Operations Research techniques in Constraint Programming, Bologna, Italy, June 2010, http://www.artist-embedded.org/artist/Overview,2022.html.
- Iuliana Bacivarov, invited talk at Efficient Execution of Kahn Process Networks on MPSoC, Mapping Applications to MPSoCs 2010, June 29-30, 2010, St. Goar, Germany, http://www.artist-embedded.org/artist/Program,1822.html.

**RWTH**
- Jovana Jovic, Simulation Challenges in the EURETILE Project, CASTNESS 2011, January 17-18, 2011, Rome, Italy
- Special session at DATE 2011: Virtual Manycore Platforms: Moving Towards 100+ Processor Cores, organized by R. Leupers and G. Martin

- R. Leupers, H. Meyr: Embedded Processor Design, Block lecture, ALARI, Lugano, Feb 2011
- ICT Technology Transfer Workshop targeting Horizon 2020, Brussels, Apr 2011, organized by R. Leupers
- S. Yakoushkin: Advanced Simulation Techniques, Joint RWTH/TU Tampere Seminar, June 2011
- R. Leupers: SoC Design Research in the UMIC Excellence Cluster, Seminar, TU Berlin, Sep 2011
- Juan Eusse: Hybrid Simulation Technology for Extensible Cores and Full System Simulation of Complex MPSoCs, Nov 2011 (Presentation at HiPEAC Computing Systems Week)
- Rainer Leupers, HiPEAC Cluster Meeting (Design and Simulation Cluster), October 2010, Barcelona, Spain
- Rainer Leupers, MPSoC Design for Wireless Multimedia, Tutorial, MIXDES, June 2010, Wroclaw, Poland
- Rainer Leupers, Cool MPSoC Design, ASCI Winter School on Embedded Systems, March 2010, Soesterburg, Netherlands
- Rainer Leupers, Design Technologies for Wireless Systems-On-Chip, Huawei ESL Symposium, September 2010, Shenzhen, People's Republic of China
- Rainer Leupers, Embedded Processor Design and Implementation, course in MSc in Embedded Systems track at ALaRI Institute, March 1-4, 2010, University of Lugano, Switzerland
- Stefan Kraemer, Advanced Simulation Techniques for Virtual Platforms, May 26, 2010, Imperial College London, London, United Kingdom
- Christoph Schumacher, Stefan Kraemer and Rainer Leupers, demonstration at DAC 2010 exhibition: parSC: parallel SystemC simulation, deterministic, accurate, fast, June 14-16, 2010, Anaheim, USA
- Christoph Schumacher, Virtual Platform Technologies for Multi-core Platforms, UMIC Day, 19 October, 2010, RWTH Aachen

**INFN**
- P. Vicini, QUonG: A GPU-based HPC System Dedicated to LQCD Computing - Symposium on Application Accelerators in High-Performance Computing, Knoxville, TN - USA, July 2011http://doi.ieeecomputersociety.org/10.1109/SAAHPC.2011.15
- D. Rossetti, Remote Direct Memory Access between NVIDIA GPUs with the APEnet 3D Torus Interconnect - SC11 - International Conference for High Performance Computing, Networking, Storage and Analysis - Seattle, WA http://nvidia.fullviewmedia.com/fb/nv-sc11/tabscontent/archive/304-wed-rossetti.html
- P.S. Paolucci, EURETILE: Brain-Inspired many-tile SW/HW Experiment, CASTNESS 2011, 17 and 18 January 2011, Rome, Italy,http://euretile.roma1.infn.it/mediawiki/img_auth.php/3/3a/EURETILE-1-PierStanislaoPaolucci.pdf
- P. Vicini, EURETILE: The HPC and Embedded Experimental HW Platform, CASTNESS 2011, 17 and 18 January 2011, Rome, Italy,http://euretile.roma1.infn.it/mediawiki/img_auth.php/6/69/EURETILE-6-PieroVicini.ppt
- R. Ammendola, apeNET+: a 3D toroidal network enabling petaFLOPS scale Lattice QCD simulations on commodity clusters, Lattice 2010, THE XXVIII INTERNATIONAL SYMPOSIUM ON LATTICE FIELD THEORY, Villasimius, Italy, June 2010, http://agenda.infn.it/contributionDisplay.py?contribId=335&sessionId=70&confId=2128
- D. Rossetti, apeNET+ Project Status, Lattice 2010, THE XXIX INTERNATIONAL SYMPOSIUM ON LATTICE FIELD THEORY, Squaw Village, Lake Tahoe, CA, USA, July 2011 - to be published on https://latt11.llnl.gov/html/proceedings.php

**Target**
- W. Geurts, G. Goossens, "Ideas for the Design of an ASIP for LQCD", CASTNESS 2011, Rome (Italy), January 17-18, 2011,http://euretile.roma1.infn.it/mediawiki/img_auth.php/3/3c/EURETILE-5-WernerGeurts.ppt
- G. Goossens, "Why Compilation Tools are the Catalyst for Multicore SoC Design", Electronic Design and Solutions Fair, Yokohama (Japan), January 27-28, 2011.
- G. Goossens, "Why Compilation Tools are a Catalyst for Multicore SoC Design", Third Friday Workshop on Designing for Embedded Parallel Computing Platforms: Architectures, Design Tools, and Applications, Design Automation and Test in Europe (DATE-2011), Grenoble (France), March 18, 2011.
- G. Goossens, P. Verbist, "Enabling the Design and Programming of Application-Specific Processors", Sophia-Antipolis Micro-Electronics Conference (SAME-2011), Sophia-Antipolis (France), October 12-13, 2011.
- G. Goossens, "Building Multicore SoCs with Application-Specific Processors", Electronic Design and Solutions Fair, Yokohama (Japan), November 16-18, 2011.

- P. Verbist, "Building software-programmable accelerators for ARM-based subsystems", ARM Technical Symposium, Taipei and Hsinchu (Taiwan), November 17-18, 2011.
- G. Goossens, "Building Multicore SoCs with Application-Specific Processors", Intl. Conf. on IP-Based SoC Design (IP-SoC-2011), Grenoble (France), December 7-8, 2011.
- G. Goossens, "Design Tools for Building Software-Programmable Accelerators in Multicore SoCs", Workshop on Tools for Embedded System Design, Sint-Michielsgestel (Netherlands), December 13, 2011.
- G. Goossens, E. Brockmeyer, W. Geurts, "Application-Specific Instruction-set Processors (ASIPs) and related design tools for tiled systems", CASTNESS 2012, Paris (France), January 26, 2012.
- G. Goossens, "How ASIP Technology can Make your RTL Blocks More Flexible", Electronic Design and Solutions Fair, Yokohama (Japan), January 28-29, 2010.
- S. Cox, G. Goossens, "Hardware Accelerator Performance in a Programmable Context: Methodology and Case Study", Embedded Systems Conference, San Jose (CA, USA), April 26-29, 2010.
- G. Goossens, "Design of Programmable Accelerators for Multicore SoCs", First Artemis Technology Conference, Budapest (Hungary), June 29-30, 2010.

**TIMA**
- Frédéric Rousseau, Requirements in Communication Synthesis for EURETILE: The use of Communication Path Formalization, CASTNESS 2012, January 26th, 2012, Paris, France
- Frédéric Rousseau, presentation of the EURETILE project in front of the Board of directors of the University Joseph Fourier, March 2011, Grenoble, France
- Frédéric Rousseau, Communication Synthesis in Low Level Software for Hierarchical Heterogeneous Systems, CASTNESS 2011, January 17-18, 2011, Rome, Italy

## 4.4. CASTNESS'11

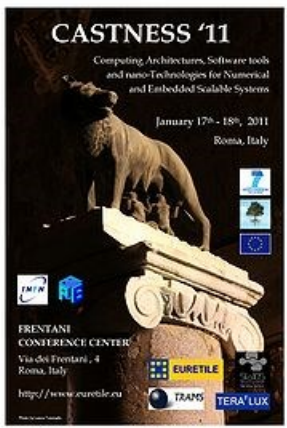

A key action conducted during this period in the framework of WP9 has been the organization of the CASTNESS'11 Workshop (Computer Architectures, Software tools and nano-Technologies for Numerical and Embedded Scalable Systems), hosted in Roma on 17-18 January 2011 at the Frentani Conference Center. http://www.congressifrentani.it

Twenty speakers, representing the four Teradevice Computing projects (EURETILE, TERAFLUX, TRAMS and SooS) presented their activities to about 60 researchers.
An archive, hosting all the presentations, is available for public access, hosted on the http://www.euretile.eu web site.

**CASTNESS'11 Agenda**

| Monday 17 January 2011: Day 1 | | | | | |
|---|---|---|---|---|---|
| **Event** | **Speaker Name** | **Affiliation** | **Duration** | **Start** | **End** |
| REGISTRATION | | | 00:40 | 13:00 | 13:40 |
| OPENING and WELCOME | Speranza Falciano | INFN Roma | 00:20 | 13:40 | 14:00 |
| SOOS: Scaling Issues in Current OS Architectures and Approaches to Overcome them | Daniel Rubio Bonilla | USTUTT-HLRS, Stuttgart | 00:20 | 14:00 | 14:20 |
| SOOS: Composition vs Concurrency | Mihai Letia | EPFL, Lausanne | 00:20 | 14:20 | 14:40 |
| SHORT BREAK | | | 00:20 | 14:40 | 15:00 |
| SOOS: Issues in Large Scale Scheduling of Distributed Applications | Tommaso Cucinotta | Scuola Superiore Sant'Anna, Pisa | 00:20 | 15:00 | 15:20 |
| SOOS: Resource Discovery and Modelling in Heterogeneous Computing Environments | Christiaan Baaij | University of Twente | 00:20 | 15:20 | 15:40 |
| SHORT BREAK | | | 00:20 | 15:40 | 16:00 |
| TRAMS: Introduction to the TRAMS Project Objectives | Antonio Rubio | UPC Barcelona | 00:20 | 16:00 | 16:20 |
| TRAMS Variability at Device Level | Andrew Brown | University of Glasgow | 00:20 | 16:20 | 16:40 |
| COFFEE BREAK | | | 00:30 | 16:40 | 17:10 |
| TRAMS: New Tools and Methods in Robust SRAM Design | Paul Zuber | Imec | 00:20 | 17:10 | 17:30 |
| TRAMS: Temperature, Voltage and Process Variations in SRAMs | Shrikanth Ganapathy | UPC Barcelona | 00:13 | 17:30 | 17:43 |
| TRAMS: Robustness of SRAM Memories | Ioana Vatajelu | UPC Barcelona | 00:13 | 17:43 | 17:56 |
| TRAMS: Carbon Nanotub Technology an Alternative in Future RAM Memories | Carmen Garcia | UPC Barcelona | 00:14 | 17:56 | 18:10 |
| SOCIAL DINNER | | | | 20:30 | |

| Tuesday 18 January 2011: Day 2 | | | | | |
|---|---|---|---|---|---|
| **Event** | **Speaker Name** | **Affiliation** | **Duration** | **Start** | **End** |
| EURETILE: Brain-Inspired many-tile SW/HW Experiment | Pier S. Paolucci | INFN Roma | 00:20 | 09:00 | 09:20 |
| EURETILE: Distributed Application Layer - Towards Seamless Programming of Many-Tiles Architectures | Iuliana Bacivarov | ETH Zurich | 00:20 | 09:20 | 09:40 |
| EURETILE: Simulation Challenges in the EURETILE Project | Jovana Jovic | RWTH Aachen | 00:20 | 09:40 | 10:00 |
| COFFEE BREAK | | | 00:30 | 10:00 | 10:30 |
| EURETILE: Communication Synthesis in Low Level Software for Hierarchical Heterogeneous Systems | Frederic Rousseau | TIMA/UJF Grenoble | 00:20 | 10:30 | 10:50 |
| EURETILE: Ideas for the design of an ASIP for LQCD | Werner Geurts | TARGET Leuven | 00:20 | 10:50 | 11:10 |
| SHORT BREAK | | | 00:20 | 11:10 | 11:30 |
| EURETILE: The HPC and Embedded Experimental HW Platform | Piero Vicini | INFN Roma | 00:20 | 11:30 | 11:50 |
| TERAFLUX: Intro | Roberto Giorgi | Università di Siena | 00:20 | 11:50 | 12:10 |
| TERAFLUX: OS | Doron Shamia | Microsoft | 00:20 | 12:10 | 12:30 |
| LUNCH | | | 01:40 | 12:30 | 14:10 |
| TERAFLUX: On Transactional Memory in TERAFLUX | Berham Khan | University of Manchester | 00:20 | 14:10 | 14:30 |
| TERAFLUX: On Computational Models | Salman Khan | University of Manchester | 00:20 | 14:30 | 14:50 |
| SHORT BREAK | | | 00:20 | 14:50 | 15:10 |
| TERAFLUX: On State-of-the-art and Plans on the Loop Nest Optimization | Konrad Trifunovic | INRIA Saclay, France | 00:20 | 15:10 | 15:30 |
| FET Plans | Jean-Marie Auger | EU FET Officer | 00:20 | 15:30 | 15:50 |
| CLOSURE | | | | 15:50 | |

# 5. APPENDIX: Euretile Project Glossary

A
- ABI - Application Binary Interface
- ALUT - Adaptive LookUp Table
- AED - Abstract Execution Device
- API - Application Programming Interface
- APEnet+ - An FPGA-based card for low latency, high bandwidth direct network interconnection based on the DNP
- ASIP - Application Specific Instruction Set Processor

B
- BER - Bit Error Rate
- BML - Byte Management Layer, framework of the OpenMPI library
- BPDLang - Bug Pattern Description Language
- BTL - Byte Transfer Layer, framework of the OpenMPI library

C
- CA - Cycle Accurate
- CDR - Clock Data Recovery
- CLI - Command Line Interface
- CQ - Completion Queue
- CRC - Cyclic Redundancy Check

D
- DAL - Distributed Application Layer
- DFM - DNP Fault Manager
- DMI - Direct Memory Interface
- DNAOS - DNA is Not just Another Operating System
- DNP - Distributed Network Processor
- DOL - Distributed Operating Layer
- DPSNN-STDP - Distributed Polychronous Spiking Neural Networks with synaptic Spiking Time Dependent Plasticity(a PSNN-STDP code natively redesigned and rewritten to exploit parallel/distributed computing systems)
- DWARF - Debugging with Attributed Record Formats, a standard format for debug information in a binary object
- DWR - DNP Watchdog Register

E
- ECC - Error correcting code
- EIR - Event-based Intermediate Representation for multi-core debugging
- EURETILE - EUropean REference TILed architecture Experiment - (all)

F
- FIT - Failure In Time
- FM - Fault manager -
- FPGA - Field-Programmable Gate Array

G
- GUI - Graphic User Interface

H
- HAL - Hardware Abstraction Layer
- HdS - Hardware dependent Software
- HFM - Host Fault Manager
- HLEM - High-level Event Monitor for multi-core debugging
- HPC - High Performance Computing
- HyNoC - Hybrid NoC simulation technology
- HySim - Hybrid Simulation technology
- HWR - Host Watchdog Register

I

- IA - Instruction Accurate
- ICE - Institute for Communication technologies and Embedded systems
- IDE - Integrated Development Environment (tools to develop and debug embedded software, integrated in a GUI)
- IMC - Interface Method Call
- INFN - Istituto Nazionale di Fisica Nucleare (National Institute for Nuclear Physics)
- I/O - Input/Output
- IOCTL - Input/Output Control, is a system call for device-specific input/output operations and other operations which cannot be expressed by regular system calls.
- IP - Intellectual Property
- IP Designer - TARGET's tool-suite for the design and programming of ASIPs
- IRISC - ICE Reduced Instruction Set Computer (RWTH's proprietary processor with that name)
- ISR - Interrupt Service Routine
- ISS - Instruction Set Simulator

J

- JIT - Just-in-time compilation

K

- KPN – Khan Process Network

L

- LDM - LiFaMa Diagnostic Message
- LiFaMa - Link Fault Manager
- LISA - Language for Instruction Set Architecture Design
- LO|FA|MO - Local Fault Monitor
- LP - Logical Process
- LQCD - Lattice Quantum-ChromoDynamics
- LSB - Least Significant Bit
- LTD - Long Term synaptic Depression
- LTL - Linear Temporal Logic
- LTP - Long Term synaptic Potentiation
- LUT - Look-Up Table
- LwIP - Lightweight IP

M

- MM - Memory Management
- MMU - Memory Management Unit
- MPI - Message Passing Interface
- MSB - Most Significant Byte
- MTL - Matching Transport Layer, framework of the OpenMPI library

N

- NIC - Network Interface Controller
- NIOS II - 32-bit microprocessor available as soft-core in Altera FPGAs (often shorthanded as "NIOS")
- nML - Not a Modelling Language (a processor architectural description language)

O

- OMPI - Open MPI, an implementation of the MPI standard.
- OPAL - Open Portability Access Layer, part of the OpenMPI library
- OpenGL - The Open Graphics Library multi-platform API for rendering computer graphics
- ORTE - Open Run-time Environment, part of the OpenMPI library
- OS - Operating System
- OSCI - Open SystemC Initiative
- OSI - Open Systems Interconnection

P

- parSC - parallel SystemC (a parallelizing SystemC execution engine)

- PCI/PCIe - Peripheral Component Interconnect - TIMA/UJF
- PCS - Physical Coding Sublayer -INFN
- PDC - Pin-Down Cache - INFN
- PEO - Process Execution Order
- PIC - Programmable Interrupt Controller
- PMA - Physical Medium Attachment
- PML - Point-to-point Message Layer, framework of the OpenMPI library
- PRBS - Pseudorandom Binary Sequence
- Presto - MPI-like library for APEnet+/DNP
- PSB - Peripheral Subsystem Bus in the VEP-EX
- PSNN - Polychronous Spiking Neural Network (A neural network which takes in consideration the delay introduced by the axonal arborization, in principle different for each synapse, and reproduces the spiking behaviour of a neural network)
- Python - A general-purpose interpreted programming language
- P2P - peer-to-peer

Q

- QUonG - LQCD on GPU platform

R

- RB - Ring Buffer
- RDMA - Remote Direct Memory Access
- RDMA GET - RDMA READ operation that implies an handshake between the sender and the receiver
- RDMA PUT - RDMA WRITE operation that implies an handshake between the sender and the receiver
- RTE - Runtime Environment - RWTH
- RTL - Register Transfer Level (also used to refer to register-transfer languages, such as VHDL or Verilog)
- RWTH - Rheinisch-Westfaelische Technische Hochschule Aachen

S

- SC - SystemC
- SCandal - SystemC Analysis for NonDeterminism Anomalies
- SCML - SystemC Modelling Language (a collection of convenience modelling objects for SystemC)
- SIMD - Single Instruction Multiple Data (also known as vector processing, a processor architectural concept to implement data-level parallelism)
- SNDNP - Service Network DNP
- SNET - Service Network
- STDP - synaptic Spiking Time Dependent Plasticity (a mechanism of synaptic evolution that depends on the relative timing between the spike incoming to a neuron, and the spike that the neuron emits. The synapses can be potentiated or depressed)
- SIP - Software Interface Protocol

T

- TCL - Tool Command Language, a general-purpose interpreted programming language
- TLM - Transaction Level Modelling
- TSBE - Target-specific Back-end for multi-core debugging

V

- V2P - Virtual to Physical
- VBE - VESA BIOS Extensions
- VEP - Virtual EURETILE Platform
- VEP-Ex - Virtual EURETILE Platform Experimental
- VHDL - VHSIC Hardware Description Language
- VLIW - Very Long Instruction Word (a processor architectural concept to implement instruction-level parallelism)

W

- WD - WatchDog

X

- XML - Extensible Markup Language
- XSD - XML Schema Definition

Ammendola, R.; Biagioni, A.; Chiodi, G.; Frezza, O.; Lo Cicero,F.; Lonardo, A.; Lunadei, R.; Paolucci, P.S.; Rossetti, D.; Salamon, A.; Salina, G.; Simula, F.; Tosoratto, L.; Vicini. P., "High-speed data transfer with FPGAs and QSFP+ modules", 2010 *JINST* 5 C12019 doi:10.1088/1748-0221/5/12/C12019

Project: **EURETILE** – European Reference Tiled Architecture Experiment
Grant Agreement no.: **247846**
Call: FP7-ICT-2009-4 Objective: FET - ICT-2009.8.1 Concurrent Tera-device Computing

**EURETILE 2010-2012 summary: first three years of activity of the European Reference Tiled Experiment.**

Pier Stanislao Paolucci, Iuliana Bacivarov, Gert Goossens, Rainer Leupers, Frédéric Rousseau, Christoph Schumacher, Lothar Thiele, Piero Vicini

The EURETILE project is funded by the European Commission through the Grant Agreement no. 247846 Call: FP7-ICT-2009-4 Objective FET-ICT-2009.8.1 Concurrent Tera-device Computing.

**www.euretile.eu**